\documentclass[conference]{IEEEtran}
\IEEEoverridecommandlockouts
\usepackage{cite}
\usepackage{amsmath,amssymb,amsfonts}
\usepackage{algorithmic}
\usepackage{graphicx}
\usepackage{textcomp}
\usepackage{xcolor}
\usepackage{listings}
\usepackage{xspace}
\usepackage{amsmath,amsthm,verbatim,amscd}
\usepackage{caption}
\usepackage{subcaption}
\usepackage[ruled, vlined]{algorithm2e}
\usepackage{booktabs, multirow}
\usepackage{tikz}
\usepackage{amssymb}
\usepackage{pifont}
\usepackage{enumitem}
\usepackage{url}
\usepackage{wrapfig}
\usepackage{balance}
\usepackage[T1]{fontenc}
\newcommand{\cmark}{\ding{51}}%
\newcommand{\xmark}{\ding{55}}%

\newtheorem{definition}{Definition}
\newcommand{\eg}{\textit{e.g.}~}
\newcommand{\ie}{\textit{i.e.}~}
\newcommand{\etal}{\textit{et al.}~}

\newcommand{\name}{\textsc{darm}\xspace}

\newcommand{\circled}[1]{\raisebox{.5pt}{\textcircled{\raisebox{-.9pt} {#1}}}} 
\def\BibTeX{{\rm B\kern-.05em{\sc i\kern-.025em b}\kern-.08em
    T\kern-.1667em\lower.7ex\hbox{E}\kern-.125emX}}

\begin{document}
\title{DARM: Control-Flow Melding for SIMT Thread Divergence Reduction - Extended Version\\
}

\author{\IEEEauthorblockN{Charitha Saumya, Kirshanthan Sundararajah, and Milind Kulkarni}
\IEEEauthorblockA{\textit{School of Electrical and Computer Engineering} \\
\textit{Purdue University}\\
West Lafayette, IN, USA \\
cgusthin@purdue.edu, ksundar@purdue.edu, milind@purdue.edu}
}

\maketitle
\thispagestyle{plain}
\pagestyle{plain}

\begin{abstract}
GPGPUs use the Single-Instruction-Multiple-Thread (SIMT) execution model where a group of threads---wavefront or warp---execute instructions in lockstep. 
When threads in a group encounter a branching instruction, not all threads in the group take the same path, a phenomenon known as control-flow divergence. 
The control-flow divergence causes performance degradation because both paths of the branch must be executed one after the other. 
Prior research has primarily addressed this issue through architectural modifications.
We observe that certain GPGPU kernels with control-flow divergence have similar control-flow structures with similar instructions on both sides of a branch. 
This structure can be exploited to reduce control-flow divergence by melding the two sides of the branch allowing threads to reconverge early, reducing divergence.
In this work, we present \name, a compiler analysis and transformation framework that can meld divergent control-flow structures with similar instruction sequences. 
We show that \name can reduce the performance degradation from control-flow divergence.
\end{abstract}

\begin{IEEEkeywords}
GPGPUs, Control-Flow Divergence, Compiler Optimizations
\end{IEEEkeywords}

\section{Introduction}
\label{sec:intro}

General Purpose Graphics Processing Units (GPGPU) are capable of executing thousands of threads in parallel, efficiently. 
%
Advancements in the programming models and compilers for GPUs have made it much easier to write data-parallel applications.
Unfortunately, exploiting data parallelism does not immediately translate to better performance.
One key reason for the lack of performance portability is that GPGPUs are not capable of executing all the  threads independently. Instead threads are grouped together
into units called \textit{warps}, and threads in a warp execute instructions in lockstep. This is commonly referred to as the Single Instruction Multiple Thread (SIMT)
execution model. 

The SIMT model suffers performance degradation when threads exhibit {\em irregularity} and can no longer execute in lockstep.
Irregularity comes in two forms, irregularity in memory accesses patterns (\ie memory divergence) and irregularity in the control-flow of
the program (\ie control-flow divergence). Memory divergence occurs when GPGPU threads needs to access memory at non-uniform locations, which
results in un-coalesced memory accesses. Un-coalesced memory accesses are bad for GPU performance because memory bandwidth can not be fully
utilized to do useful work.

Control-flow divergence occurs when threads in a warp diverge at branch instructions.
At the {\em diverging} branch, lockstep execution can not be maintained because threads in a warp may want to execute different basic bocks (\ie diverge).
%
%
Instead, when executing instructions along a diverged path, GPGPUs mask out the threads that do not want to take that path. The threads {\em reconverge} at the Immediate Post-DOMinator (IPDOM) of a divergent branch---the instruction that all threads from both branches want to execute. This style of IPDOM-based reconvergence is implemented in hardware in most GPGPU architectures to maintain SIMT execution.
Even though IPDOM-based reconvergence can handle arbitrary control-flow, it imposes a significant performance penalty if a program has
a lot of divergent branches. In the IPDOM reconvergence model, instructions executed on divergent branches necessarily cannot utilize the full width of a SIMD unit. If the code has a lot of nested divergent branches or divergent branches inside loops, this style of execution causes significant under-utilization
of SIMD resources. 

For some GPGPU applications divergent branches are unavoidable, and there have been many techniques proposed to address this issue both in hardware and software.
Proposals such as Dynamic warp formation~\cite{dynamicwarp}, Thread block compaction~\cite{compact_thblk} and Dual-path execution~\cite{dualpath}
focus on mitigating the problem at the hardware level by changing how threads are scheduled for execution and making sure that
threads following the same path are grouped together. Unfortunately, such approaches are not useful on commodity GPGPUs.

\begin{table}[tb]
  \centering
  \caption{Comparison of techniques for divergence reduction}
  \begin{tabular}{|c|c|c|c|}
  \hline
  \multirow{2}{*}{\textbf{\begin{tabular}[c]{@{}c@{}}Control-flow and instruction\\ Pattern\end{tabular}}} & \multicolumn{3}{c|}{\textbf{Technique}}                                                                                   \\ \cline{2-4} 
                                                                                                           & \begin{tabular}[c]{@{}c@{}}Tail \\ Merging\end{tabular} & \begin{tabular}[c]{@{}c@{}}Branch \\ Fusion\end{tabular} & \name \\ \hline
  \begin{tabular}[c]{@{}c@{}}Diamond control-flow with \\ identical instruction sequences\end{tabular}     & \cmark                                                        & \cmark                                                         & \cmark      \\ \hline
  \begin{tabular}[c]{@{}c@{}}Diamond control-flow with \\ distinct instruction sequences\end{tabular}      & \xmark                                                        & \cmark                                                         & \cmark      \\ \hline
  Complex control-flow                                                                                     & \xmark                                                        & \xmark                                                         & \cmark     \\ \hline
  \end{tabular}
  \label{tab:related_work}
  \vspace{-2.0em}
\end{table}

There have also been efforts to reduce divergence through compiler approaches that leverage the observation that different control-flow paths often contain similar instruction (sub)sequences. {\em Tail merging}~\cite{gen_tail_merge_sas03} identifies branches that have identical sequences of code and introduces early jumps to {\em merged} basic blocks, with the effect of reducing divergence. {\em Branch fusion} generalizes tail merging to work with instruction sequences that may not be identical~\cite{branch_fusion}. 
However, branch fusion cannot analyze complex control-flow and hence it is restricted to simple {\em if-then-else} branches where each path has a single basic block (\ie diamond-shaped control-flow).

This paper introduces a more general, software-only approach of exploiting similarity in divergent paths, called {\em control-flow melding}. 
 Control-flow melding is a general control-flow transformation which can meld similar
control-flow \textit{subgraphs} inside a {\em if-then-else} region (not just individual basic blocks). 
By working hierarchically, recursively melding divergent control-flow at the level of subgraphs of the CFG, control-flow melding can handle substantially more general control structures than prior work. This paper describes \name, a realization of control-flow melding for general GPGPU programs. Table~\ref{tab:related_work} compares the capabilities of \name with branch fusion and tail merging.

\name works in several steps. First, it detects divergent {\em if-then-else} regions and splits the divergent regions into Single Entry Single Exit (SESE) control-flow subgraphs. Next it uses a hierarchical sequence alignment technique to \textit{meld} profitable
control-flow subgraphs, repeatedly finding subgraphs whose control-flow structures and constituent instructions can be aligned.
Once a fixpoint is reached, \name uses this hierarchical alignment to generate code for the region with reduced control-flow divergence. 

The main contributions of the paper are,
\begin{itemize}[leftmargin=*]
  \item \textbf{D}ivergence-\textbf{A}ware-\textbf{R}egion-\textbf{M}elder (\name), a realization of control-flow melding that identifies profitable melding opportunities in divergent {\em if-then-else}
        regions of the control-flow using a hierarchical sequence alignment approach and then melds these regions to reduce control-flow divergence.
  \item An implementation of \name  in LLVM~\cite{llvm} that can be applied to GPGPU programs written in HIP~\cite{HIP} or CUDA~\cite{cuda}. Our implementation of \name is publicly available as an archival repository\footnote{https://doi.org/10.5281/zenodo.5784768} and up-to-date version is available in GitHub\footnote{https://github.com/charitha22/cgo22ae-darm-code}.
  \item An evaluation of \name on a set of synthetic GPU programs and a set of real-world GPU applications showing its effectiveness
\end{itemize}

\section{Background}
\label{sec:background}
\subsection{GPGPU Architecture}
Modern GPGPUs have multiple processing cores, each of which contains multiple parallel lanes (\ie SIMD units), a vector register file and 
a chunk of shared memory. The unit of execution is called a warp (or wavefront). A warp is a collection of threads executed in lock-step on 
a SIMD unit. Shared memory is shared among the warps executing on a core.
A branch unit takes care of control-flow divergence by maintaining a SIMT stack to enforce IPDOM based reconvergence, as discussed in Section~\ref{sec:intro}.
GPGPU programming abstractions like CUDA~\cite{cuda} or HIP~\cite{HIP} gives the illusion of data parallelism with independent threads. 
However, during real execution, a group of program instances (\ie threads) are mapped to a warp and executed in lock-step. Therefore control-flow divergence in 
SPMD programs is detrimental to the performance because of the SIMT execution limitations.

\subsection{LLVM SSA form and GPU Divergence Analysis}
LLVM~\cite{llvm} is a general framework for building compilers, optimizations and code generators. Most of the widely adopted GPGPU compilers~\cite{nvcc,HIPCC} are built on 
top of the LLVM infrastructure.
 LLVM uses a target-independent intermediate representation, LLVM-IR, that enables 
implementing portable compiler optimizations. LLVM-IR uses static single assignment form~\cite{ssa_form} which requires that every program variable is assigned once and is defined before being used. 
SSA form uses $\phi$ nodes to resolve data-flow when branches are present, selecting which definition should be chosen at a confluence of different paths.
In GPGPU compilers, a key step in identifying divergent control-flow regions is performing compiler analyses to identify divergent variables (or branches)~\cite{llvm_div_analysis,branch_fusion}.
A branch is divergent if the branching condition evaluates to a non-uniform value for different threads in a warp.
If the branching condition is divergent,
threads in a warp will have to take different control-flow paths at this point. 
LLVM's divergence analysis tags a branch as divergent, if the branching condition is either data-dependent 
or sync-dependent on a divergent variable (such as thread ID)~\cite{llvm_div_analysis}, though more sophisticated divergence 
analyses have been proposed~\cite{divanalysispopl2021}.
\section{Motivating Example}
\label{sec:overview}

\begin{figure}[tb]
  \begin{lstlisting}[language=c, basicstyle=\small, frame=single]
  __global__ static void bitonicSort(int *values) {
    // copy data from global memory to shared memory
    __syncthreads();
    for (unsigned int k = 2; k <= NUM; k *= 2) {
      for (unsigned int j = k / 2; j > 0; j /= 2) {
        unsigned int ixj = tid ^ j;
        if (ixj > tid) {
          if ((tid & k) == 0) {
            if (shared[ixj] < shared[tid])
              swap(shared[tid], shared[ixj]);
          }
          else {
            if ( shared[ixj] > shared[tid])
              swap(shared[tid], shared[ixj]);
          }
        }
        __syncthreads();
      }
    } // write data back to global memory
  }
  \end{lstlisting}
  \vspace{-0.5em}
  \caption{Bitonic sort kernel}
  \vspace{-2.0em}
  \label{fig:bitonicsort_kernel}
\end{figure}
Bitonic sort is a kernel used in many parallel sorting algorithms such as bitonic merge sort and Cederman's quicksort~\cite{bitonic, gpu_quicksort_cederman}.
Figure~\ref{fig:bitonicsort_kernel} shows a CUDA implementation of bitonic sort. 
This kernel is our running example for describing \name's control-flow melding algorithm. 

In this kernel, the branch condition at line 8 depends on the {\em thread ID}. Therefore it is divergent. 
Since the divergent branch is located inside a loop, the execution of the two sides of the branch needs to be serialized many times, 
resulting in high control-flow divergence.
However the code inside the {\em if} (line 9-10) and {\em else} (line 13-14) sections of the divergent branch are similar in two ways. 
First, both code sections have the same control-flow structure (\ie {\em if-then} branch). Second, instructions along the two paths are
also similar. Both conditions compare two elements in the {\em shared} array and perform a {\em swap} operation. 
Therefore the contents of the {\em if} and {\em else} sections can be melded to reduce control-flow divergence. 
Both code sections consists of shared memory loads and store operations. In the unmelded version of the code these shared memory 
operations will have to be serialized due to thread-divergence. However, if the two sections are melded threads can issue the memory instructions
in the same cycle resulting in improved performance. 

Existing compiler optimizations such as tail merging and branch fusion cannot be applied to this case.
Tail merging is applicable only if two basic blocks have a common destination and have identical instruction sequences at their tails.
However in bitonic sort, the {\em if} and {\em then} sections of the divergent branch have multiple basic blocks, and the compiler cannot apply
tail merging. Similarly branch fusion requires diamond shaped control-flow and does not work if the {\em if} and {\em else}
sections of the branch contain complex control-flow structures. 

\name solves this problem in two phases. 
In the analysis phase (Section~\ref{sec:design:melding_profitability}), 
\name analyzes the control-flow region dominated by a divergent branch to find isomorphic sub-regions that are in the true and false paths of the divergent branch. These isomorphic sub-region pairs are aligned based on their 
melding profitability using a sequence alignment strategy. 
Melding profitability is a compile-time approximation of the percentage of thread cycles that can be saved by melding two control-flow regions. 
Next, \name choses profitable sub-region pairs in the alignment (using a threshold) and computes an instruction alignment for corresponding basic blocks in the two regions. 
In the code generation phase (Section~\ref{sec:design:codegen}), \name uses this instruction alignment to meld corresponding basic blocks in the sub-region pair. 
This melding is applied iteratively until no further profitable melding can be performed. 
\name's melding transformation is done in SSA form, therefore the resulting CFG can be optimized further using other compiler optimizations (Sections~\ref{sec:design:unpred} and~\ref{sec:design:pre_and_post}).

\section{Detailed Design}
\label{sec:design}
In this section we describe the algorithm used by \name to meld similar control-flow subgraphs. 
First we define the following terms used in our 
algorithm description.
\subsection{Preliminaries and Definitions}
\begin{definition}
  \textbf{Simple Region} : A simple region is a subgraph of a program's CFG that is connected to the remaining CFG with only two edges, an entry edge and an exit edge. 
\end{definition}
\begin{definition}
  \textbf{Region} : A region of the CFG is characterized by two basic blocks, its entry and exit. All the basic blocks inside a region are dominated by its entry and post-dominated by its exit. Region with entry $E$ and 
  exit $X$ is denoted by the tuple $(E,X)$. LLVM regions are defined similarly~\cite{llvm_region,structure_tree}. 
\end{definition}
\begin{definition}
  \label{def:sese_subgraphs}
  \textbf{Single Entry Single Exit Subgraph} : Single entry single exit (SESE) subgraph is either a simple region or a single basic block with a single predecessor and 
  a successor.
\end{definition}

\noindent
Note that a region with entry $E$ and exit $X$ can be transformed into a simple region by introducing a new entry and exit blocks $E_{new}$, $X_{new}$. 
All successors of $E$ are moved to $E_{new}$ and $E_{new}$ is made the single successor of $E$. Similarly, all predecessors of $X$ are moved to $X_{new}$ and 
a single exit edge is added from $X_{new}$ to $X$. 
\begin{definition}
  \textbf{Simplified Region} : A region with all its subregions transformed into simple regions is called a simplified region.
\end{definition}
We now turn to the steps the \name compiler pass takes to reduce control divergent code.

\subsection{Detecting Meldable Divergent Regions}

First \name needs to detect divergent branches in the CFG.
We use LLVM's built-in divergence analysis to decide if a branch is divergent or not (Section~\ref{sec:background}).
The smallest CFG region enclosing a divergent branch is called the {\em divergent region} corresponding to this branch.
Melding transformation is applied only to divergent regions of the CFG. The next step is to decide if a divergent region contains control-flow subgraphs
(definition~\ref{def:sese_subgraphs}) 
that can be safely melded.  
\begin{definition}
  \label{def:meld_div_region}
  \textbf{Meldable Divergent Region}: A simplified region $R$ with entry $E$ and exit $X$ is said to be meldable and divergent if the following conditions are met,
  \begin{enumerate}[leftmargin=*]
    \item The entry block of $R$ has a divergent branch
    \item Let $B_T$ and $B_F$ be the successor blocks of $E$. $B_T$ does not post-dominate $B_F$ and $B_F$ does not post-dominate $B_T$
  \end{enumerate}
\end{definition}

\noindent
According to definition~\ref{def:meld_div_region}, a {\em meldable divergent region} has a divergent branch at its entry (condition 1). This makes sure that our melding transformation
is only applied to divergent regions, and non-divergent parts of the control-flow are left untouched. Condition 2 ensures
that paths $B_T \rightarrow X$ (\ie true path) and $B_F \rightarrow X$ (\ie false path) consists of at least one SESE subgraph and these subgraphs from the two paths can potentially be melded to reduce 
control-flow divergence. Consider our running example in Figure~\ref{fig:bitonicsort_kernel}. When this kernel is compiled with {\em ROCm HIPCC} GPU compiler
~\cite{HIP} with {\em -O3}
optimization level into LLVM-IR, we get the CFG shown in Figure~\ref{fig:melding_flow_a}. Note that the compiler aggressively unrolls both the loops (lines 4 and 5) 
in the kernel, and the resulting CFG consists of multiple repeated segments of the inner loop's body (lines 6-17). In Figure~\ref{fig:melding_flow_a}, only one unrolled 
instantiation of the loop body is shown. As explained in Section~\ref{sec:overview}, this kernel contains a divergent branch, which is at the end of basic block $\%B$.
Also $\%B$'s two successors $\%C$ and $\%D$ do not post-dominate each other. Therefore the region $(\%B,\%G)$ is a meldable divergent region.

\subsection{Computing Melding Profitability}
\label{sec:design:melding_profitability}

Definition~\ref{def:meld_div_region} only allows us to detect regions that may contain meldable control-flow subgraphs. It does not tell us whether it is legal to meld them or melding them will improve performance.
First we need to define what conditions needs to be satisfied for two SESE subgraphs to be meldable. 
\begin{definition}
  \label{def:meldable_sese}
  \textbf{Meldable SESE Subgraphs}: SESE subgraphs $S1$ and $S2$ where $S1$ belongs to the true path and $S2$ belongs to the false path are meldable if any 
  one of the following conditions are satisfied,
  \begin{enumerate}[leftmargin=*]
    \item Both $S1$ and $S2$ have more than one basic block and they are structurally similar \ie isomorphic.
    \item $S1$ is a simple region and $S2$ consists of a single basic block or vice versa.
    \item Both $S1$ and $S2$ consists of single basic block.
  \end{enumerate}
\end{definition}
\begin{figure}[!t]
  \centering
  \includegraphics[width=0.48\textwidth,height=0.3\textwidth]{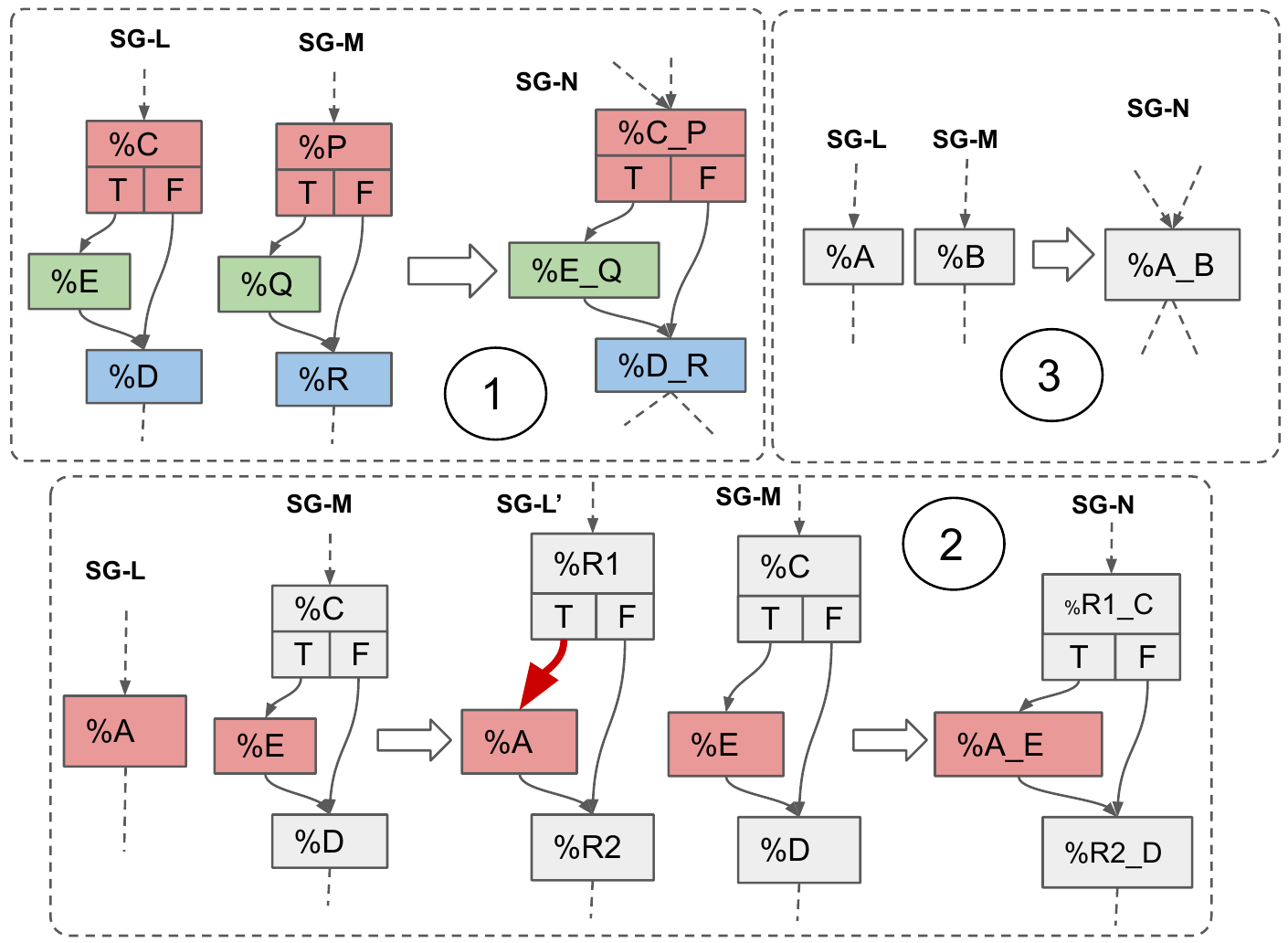}
  \caption{Examples showing the 3 cases considered by \name to detect meldable subgraphs}
  \label{fig:sese_melding_cases}
  \vspace{-1.5em}
\end{figure}
\noindent
Definition~\ref{def:meldable_sese} ensures that any two SESE subgraphs that meets any one of these conditions can be melded without introducing additional divergence 
to the control-flow. Note that we do not consider subgraphs that contain {\em warp-level intrinsics}\cite{warp_primitives} for melding because melding such subgraphs
can cause deadlock. 
Figure~\ref{fig:sese_melding_cases} shows three examples where each of the above conditions are applicable. 
Assume in each example subgraphs $L$ and $M$ are in a divergent region $(E,X)$ and only one of the subgraphs are executed from any program path
from $E$ to $X$. (\ie any 
thread in warp that executes $E$ must either go through $L$ or $M$ but not both).

  \textbf{Region-Region Melding :} 
  In case \circled{1},  two SESE subgraphs $L$ and $M$ 
  are isomorphic, therefore they can be melded to have the same control-flow structure (subgraph $N$ in Figure~\ref{fig:sese_melding_cases}-\circled{1}). 
   In the melded subgraph $N$, basic blocks $\%C\_P$ and $\%D\_R$ are 
  guaranteed to post-dominate $E$ and threads can reconverge at these points resulting in reduction in control-flow divergence. Also the structural similarity in 
  case \circled{1} ensures that we do not introduce any additional branches into the melded subgraph.
  
  \textbf{Basic block-Region Melding :}
  In case \circled{2}, basic block $\%A$ (in subgraph $L$) can potentially be melded with any basic block in CFG $M$. Assume that basic blocks $\%A$ and
  $\%E$ have the most melding profitability (melding profitability described later). First we replicate the control-flow structure of $M$ to create a new CFG $L'$. Then 
  we place $\%A$ in $L'$ such that $\%A$ and $\%E$ are in similar positions in the the two CFGs $L'$ and $M$. We also ensure the correctness of the 
  program by concretizing the branch conditions in $L'$ to always execute $\%A$ and create $\phi$ nodes at dominance frontiers of $\%A$ to make sure values defined inside $\%A$ are reached to their users~\cite{ssa_form}. In this example branch at end of basic block $\%R1$ will always take the edge $\%R1-\%A$ (bold arrow in subgraph $L'$) and $\phi$ nodes will be added to $\%R2$. Now subgraphs $L'$ and $M$ are isomorphic and therefore can be melded similar to case \circled{1}. We refer to this process as {\em Region Replication}. Main benefit of region replication is that it allows us to
  meld $\%A$ with any profitable basic block in subgraph $M$ and resultant subgraph $N$ has less divergence because threads can reconverge at basic blocks $\%R1\_C$ and $\%R2\_D$ in melded subgraph $N$. 
  
  \textbf{Basic block-Basic block Melding :}
  Case \circled{3} is the simplest form where two SESE basic blocks are melded. 

A meldable divergent region can potentially have multiple SESE subgraphs in its true and false paths. Therefore we need a strategy to figure out which subgraph pairs to 
meld. We formulate this as a sequence alignment problem as follows. First, we obtain a ordered sequence of subgraphs in true path and false of the divergent region. 
Subgraphs are ordered using the post-dominance relation of their entry and exit blocks. For example, if entry node of subgraph $S_2$ post-dominates exit node of 
subgraph $S1$, then $S2$ comes after $S1$ in the order and denoted as $S1 \prec S2$.
A subgraph alignment is defined as follows,
\begin{definition}
  \label{def:subgraph_align}
  \textbf{Subgraph Alignment}: Assume a divergent region $(E,X)$ has ordered SESE subgraphs $\{S_1^T,S_2^T,\dots,S_m^T\}$ in its true path and ordered subgraphs $\{S_1^F,S_2^F,\dots,S_n^F\}$ 
  in the false path. A subgraph alignment is an ordered sequence of tuples $A = \{(S_{i0}^T,S_{j0}^F),(S_{i1}^T,S_{j1}^F),\dots,(S_{ik}^T,S_{jk}^F)\}$ where,
  \begin{enumerate}
    \item if $(S_{p}^T,S_{q}^F) \in A$ then $S_p^T$ and $S_q^F$ are meldable subgraphs 
    \item if $(S_{p1}^T,S_{q1}^F) \prec (S_{p2}^T,S_{q2}^F)$ then $S_{p1}^T \prec S_{p2}^T$ and $S_{q1}^T \prec S_{q2}^T$
  \end{enumerate}
\end{definition}

\noindent
According to definition~\ref{def:subgraph_align}, only meldable subgraphs are allowed in a alignment tuple and if the aligned subgraphs are melded, the resultant 
control-flow graph does not break the original dominance and post-dominance relations of the subgraphs. 

Given a suitable alignment scoring function $F$
 and gap penalty function $W$, we can find an optimal subgraph alignment using a sequence alignment method such as Smith-Waterman~\cite{smith_waterman} algorithm.
The scoring function $F$ measures the profitability of melding two meldable subgraphs $S1$ and $S2$. 
Prior techniques have employed instruction frequency to approximate the profit of merging two functions\cite{rocha19, rocha20}.
We use a similar method to define subgraph melding profitability.
First we define the melding profitability of two basic blocks $b1$ and $b2$ as follows,
{\small $$MP_B(b1,b2) = \frac{\sum_{i \in Q}min(freq(i,b1),freq(i,b2)) \times w_i}{lat(b1)+lat(b2)}$$ }

\noindent
Here $Q$ is set of all possible instruction types available in the instruction set (\ie LLVM-IR opcodes). $lat(b)$ is the static latency of 
basic block which can be calculated 
by summing the latencies of all instructions in $b$. $w_i$ is the latency of instruction type $i$. The idea here is to approximate the percentage of instruction
 cycles that can be saved
by melding the instructions in $b1$ and $b2$ assuming a best-case scenario (\ie all common instructions in $b1$ and $b2$ are melded regardless of their order). For example, two basic blocks with identical opcode frequency profile will have a profitability value 0.5. 

Because meldable subgraphs are isomorphic, 
there is a one-to-one mapping between basic blocks (\ie corresponding basic blocks). 
For example, in Figure~\ref{fig:sese_melding_cases} case \circled{1} the basic block mapping for CFGs $L$ and $M$ are $\{(\%C,\%P),(\%E,\%Q),(\%D,\%R)\}$. Assume the mapping of basic blocks in $S1$ and $S2$ is denoted by $O$.
Subgraph melding profitability $MP_S$ of subgraphs $S1$ and $S2$ is defined in terms of melding profitabilities of their corresponding basic blocks.
  {\small $$MP_S(S1, S2) = \frac{\sum_{(b1,b2) \in O} MP_B(b1,b2) \times (lat(b1)+lat(b2))}{\sum_{(b1,b2) \in O} lat(b1)+lat(b2)}$$ }

Similar to $MP_B$, $MP_S$ measures the percentage of instruction cycles saved by melding two SESE subgraphs. This metric is an over-approximation,
however it provides a fast way of measure the melding profitability of two subgraphs that works well in practice. 
We use $MP_S$ as the scoring function for subgraph alignment.

\noindent
\textbf{Instruction Alignment}: Notice that our subgraph melding profitability metric (\ie $MP_S$) prioritizes subgraph pairs that have many similar instructions in their corresponding 
basic blocks. Therefore when melding two corresponding basic blocks we must ensure that maximum number of similar instructions are melded together. 
This requires computing an alignment of two instruction sequences such that if they are melded using this alignment, the number of instruction
cycles saved will be maximal.
We use the approach used in Branch Fusion~\cite{branch_fusion} to compute an optimal alignment for two instructions sequences.
In this approach compatible instructions are aligned together and instructions with higher latency are prioritized to be aligned  over lower latency instructions. Compatibility of two instructions for melding depends on 
a number of conditions like having the same opcode and 
types of the operands being compatible. We used the criteria described by Rocha et al.~\cite{rocha20} to determine this compatibility.
This instruction alignment model uses a gap penalty for unaligned instructions because extra branches needs to be generated to
conditionally execute these unaligned instructions. 
Our melding algorithm does not depend on the sequence alignment algorithm used for instruction alignment computation.
We use Smith-Waterman algorithm~\cite{smith_waterman} to compute the instruction alignment because prior work~\cite{branch_fusion} has shown its effectiveness.
\begin{figure*}[ht]
  \centering
  \begin{subfigure}[b]{0.35\textwidth}
    \centering
    \includegraphics[width=1.0\textwidth,height=0.17\textheight]{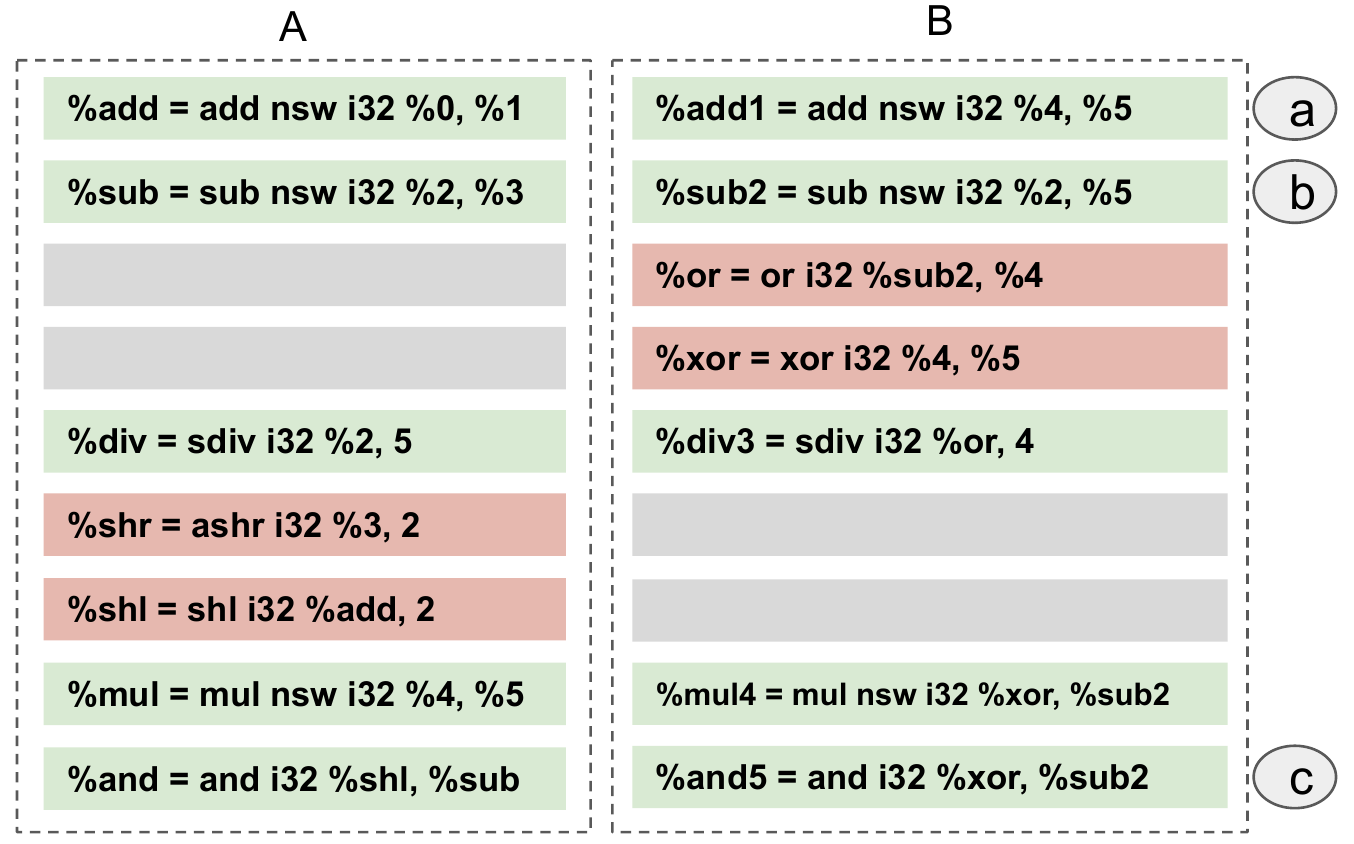}
    \caption{}
    \label{fig:instr_align_example}
  \end{subfigure}
  \hfill
  \begin{subfigure}[b]{0.25\textwidth}
    \centering
    \includegraphics[width=1.0\textwidth,height=0.16\textheight]{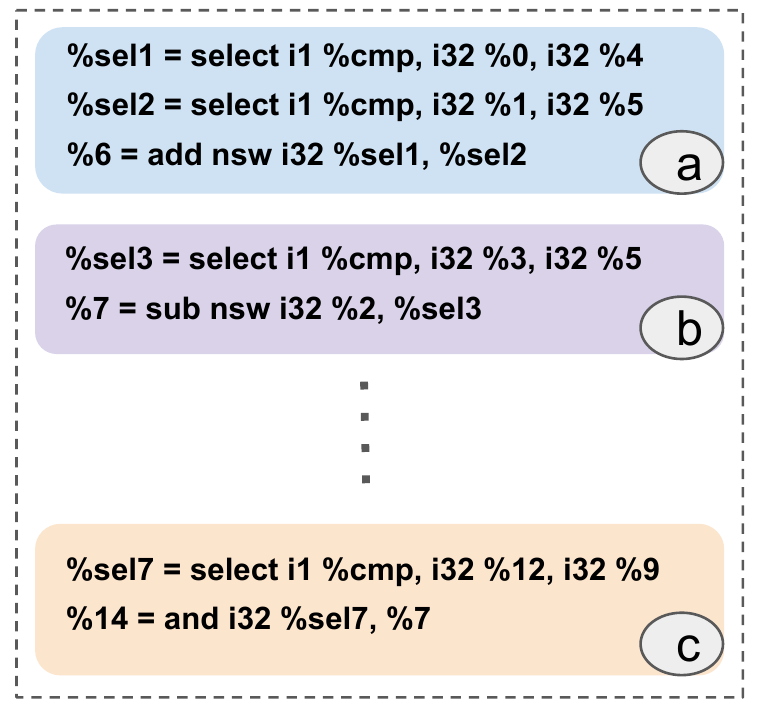}
    \caption{}
    \label{fig:instr_melding_example}
  \end{subfigure}
  \hfill
  \begin{subfigure}[b]{0.25\textwidth}
    \centering
    \includegraphics[width=1.0\textwidth,height=0.17\textheight]{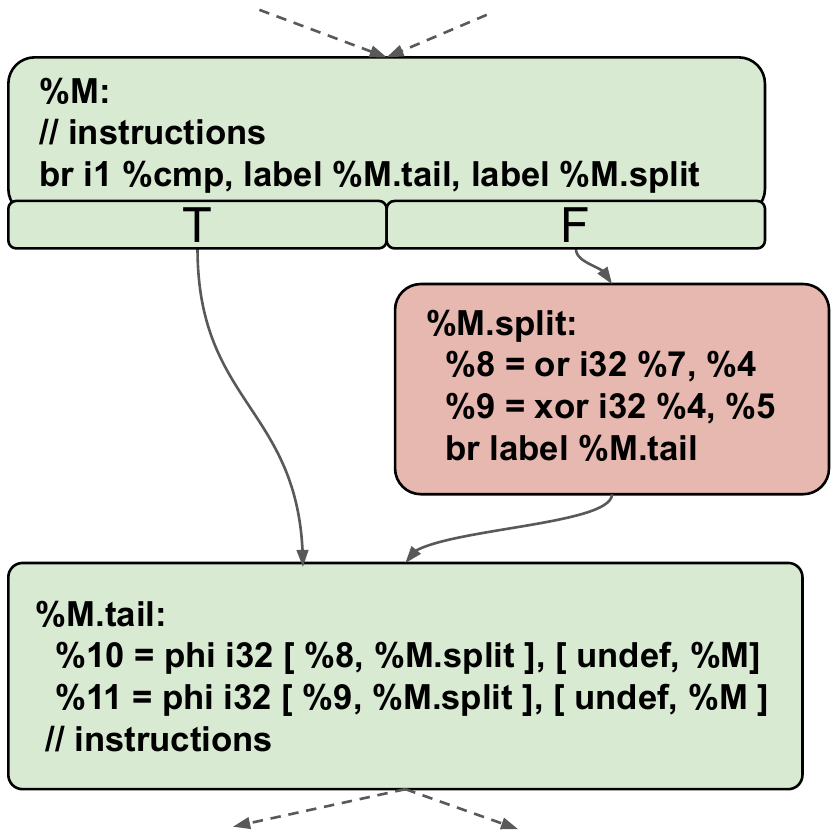}
    \caption{}
    \label{fig:unpredication_example}
  \end{subfigure}
  \caption{ 
    (a) Instruction alignment result for two basic blocks $A$ and $B$, 
    (b) Code generated by \name for aligned instructions \circled{a}, \circled{b} and \circled{c} in Figure~\ref{fig:instr_align_example}, 
    (c) Unpredication applied to the unaligned instructions of basic block $B$ in figure~\ref{fig:instr_align_example} 
  }
  \vspace{-1em} 
\end{figure*}
Figure~\ref{fig:instr_align_example} shows the instruction alignment computed for two basic blocks $A$ and $B$. Aligned instructions are 
shown in green and instructions aligned with a gap are in red.

\subsection{\name Code Generation}
\label{sec:design:codegen}
\setlength{\textfloatsep}{0pt}
\setlength{\floatsep}{0pt}
\begin{algorithm}[!htb]
  \SetAlgoLined
  \SetKwRepeat{Do}{do}{while}
  \KwIn{SPMD function $F$}
  \KwOut{Melded SPMD function $F_{out}$}
  \Do{changed}
  {
    changed $\leftarrow$ false \\
    \For{BB in F}
    {
      R, C $\leftarrow$ GetRegionFor(BB) \\
      \If{IsMeldableDivergent(R)}
      {
        SimplifyRegion(R) \\
        A $\leftarrow$ ComputeSubgraphAlignment(R) \\
        \For{($S_T,S_F$, profit) in A}
        {
          \If{profit $\geq$ threshold}
          {
            Meld($S_T,S_F$, C) \\
            changed $\leftarrow$ true 
          }
        }
      }
      \If{changed}
      {
        SimplifyFunction(F) \\
        RecomputeControlFlowAnalyses(F) \\
        break
      }
    }
  }
  \caption{\name Algorithm}
  \label{algo:main}
\end{algorithm}
\begin{algorithm}[!htb]
  \SetAlgoLined
  \KwIn{SESE subgraphs $S_T$,$S_F$, Condition C}
  \KwOut{Melded SESE subgraph $S_{out}$}
  \caption{SESE Subgraph melding Algorithm}
  List blockPairs $\leftarrow$ Linearize($S_T,S_F$) \\
  List A $\leftarrow$ empty \\
  \For{($B_T,B_F$) in blockPairs}
  {
    List instrPairs $\leftarrow$ ComputeInstrAlignment($B_T,B_F$) \\
    A.append(instrPairs)
  }
  PreProcess($S_T,S_F$) \\ 
  Map operandMap $\leftarrow$ empty \\
  \For{$P$ in A}
  {
    $I_{melded}$ $\leftarrow$ Clone($P$) \\
    Update(operandMap, $I_{melded}$, $P$) \\
  }
  \For{$P$ in A}
  {
    SetOperands($P$, operandMap, C) 
  }
  RunUnpredication() \\
  RunPostOptimizations()
  \label{algo:subgraph_melding}
\end{algorithm}

\name's control-flow melding procedure is shown in algorithm~\ref{algo:main}. This algorithm takes in a SPMD function $F$ and iterates over
all basic blocks in $F$ to check if the basic block is an entry to a meldable divergent region ($R$) according to the conditions in Definition~\ref{def:meld_div_region}. 
We use $\mathit{Simplify}$ to convert all subregions inside $R$ in to simple regions.

We compute the optimal subgraph alignment for the two sequences of subgraphs in the true and false paths of $R$.
We meld each subgraph pair in the alignment if the melding profitability is greater than some threshold.  
Subgraph melding changes the control-flow of
$F$. Therefore we first simplify the control-flow (using LLVM's {\em simplifycfg}) and then 
recompute the control-flow analyses (\eg dominator, post-dominator and region tree) required for the melding pass. We apply the melding procedure
on $F$ again until no profitable melds can be performed.

Algorithm~\ref{algo:subgraph_melding} shows the procedure for melding two subgraphs $S_T$ and $S_F$.
$C$ is the branching condition of the meldable divergent region containing $S_T$ and $S_F$.
First the two subgraphs are linearized in pre-order to form a list of corresponding basic block pairs. Processing the basic blocks in pre-order 
ensures that dominating definitions are melded before their uses. For each basic block pair in this list we compute  
an optimal alignment of instructions.
Each pair in the alignment falls into two categories, {\em I-I} and {\em I-G}. I-I is a proper alignment with two instructions 
and I-G is an instruction aligned with a gap. Our alignment makes sure that in a match the two instructions are always meldable into one instruction (\eg a {\em load}
is not allowed to align with a {\em store}). 
First we traverse the alignment pair list and clone the aligned instructions. For I-I pairs, we clone a single instruction because they can be melded. 
During cloning, we also update the $\mathit{operandMap}$, which maintains a mapping between aligned and melded LLVM values.
We perform a second pass over the instruction alignment to set the operands of cloned instructions ($\mathit{SetOperands}$).
Assume we are processing an I-I pair with instructions $I_T,I_F$ and cloned instruction is $I_{melded}$. For each operand of $I_{melded}$,
the corresponding operands from 
$I_T$ and $I_F$ are looked up in $operandMap$ because an operand might be an already melded instruction. 
If the resultant two operands from $I_T$ and $I_F$ are the same, we just use that value as the operand. If they are different, we generate a {\em select} instruction
to pick the correct operand conditioned by $C$. For an I-G pair, operands are first looked up in $operandMap$ and the result is copied to $I_{melded}$.
Consider the instruction alignment in figure~\ref{fig:instr_align_example}. Figure~\ref{fig:instr_melding_example} shows the generated code 
for aligned instruction pairs \circled{a}, \circled{b} and \circled{c}. In case \circled{a}, two select instructions are needed because both operands maps to 
different values ($\%0$, $\%4$ and $\%1$, $\%5$). In case \circled{b}, the first operand is the same ($\%2$) for both instructions, therefore only one select is needed. 
In case \circled{c}, both first and second operands are different for the two instructions. However the second operands map to same melded instruction $\%7$, so only 
one select is needed. Note that $\%cmp$ is the branching condition for the divergent region, and we use that for selecting the operands. \\

\begin{figure*}[ht]
  \centering
  \begin{subfigure}[b]{0.12\textwidth}
      \includegraphics[width=\textwidth]{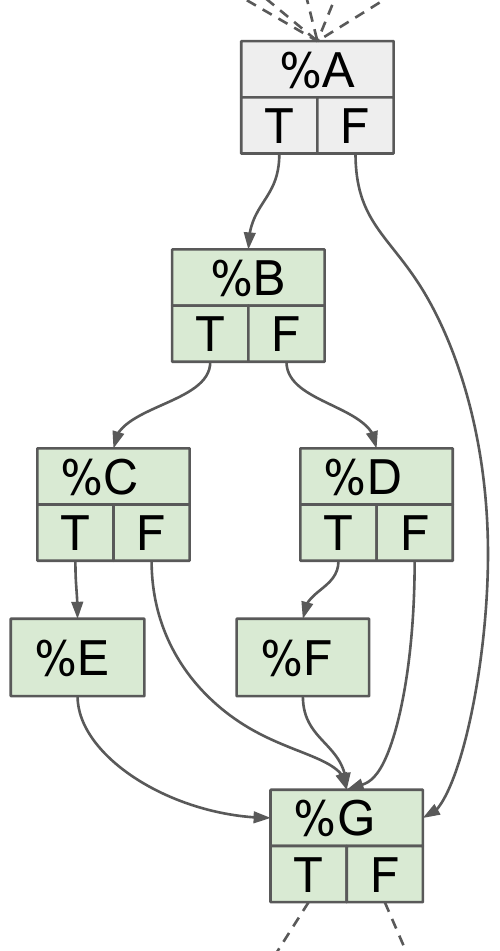}
      \caption{}
      \label{fig:melding_flow_a}
  \end{subfigure}
  \hfill
  \begin{subfigure}[b]{0.11\textwidth}
      \includegraphics[width=\textwidth]{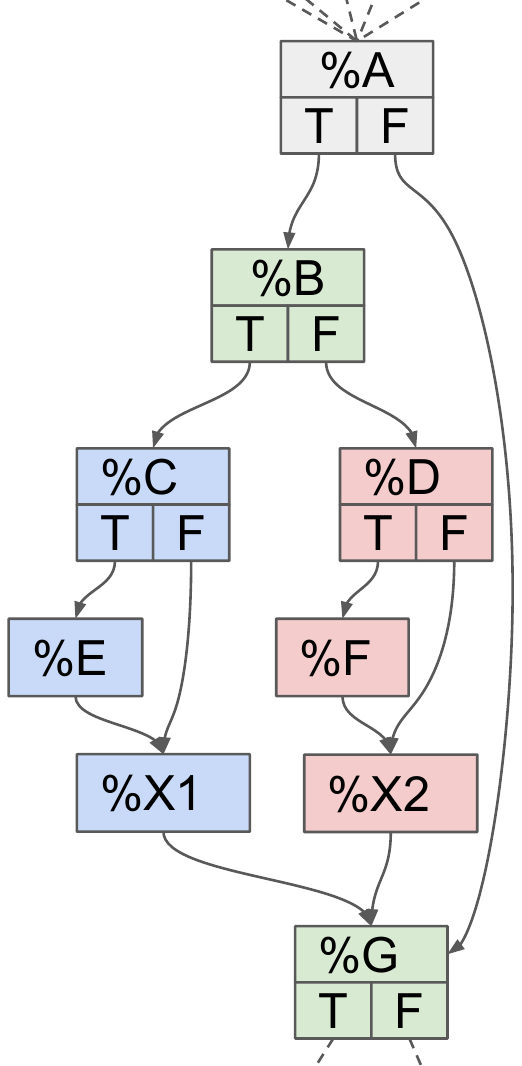}
      \caption{}
      \label{fig:melding_flow_b}
  \end{subfigure}
  \hfill
  \begin{subfigure}[b]{0.095\textwidth}
      \includegraphics[width=\textwidth]{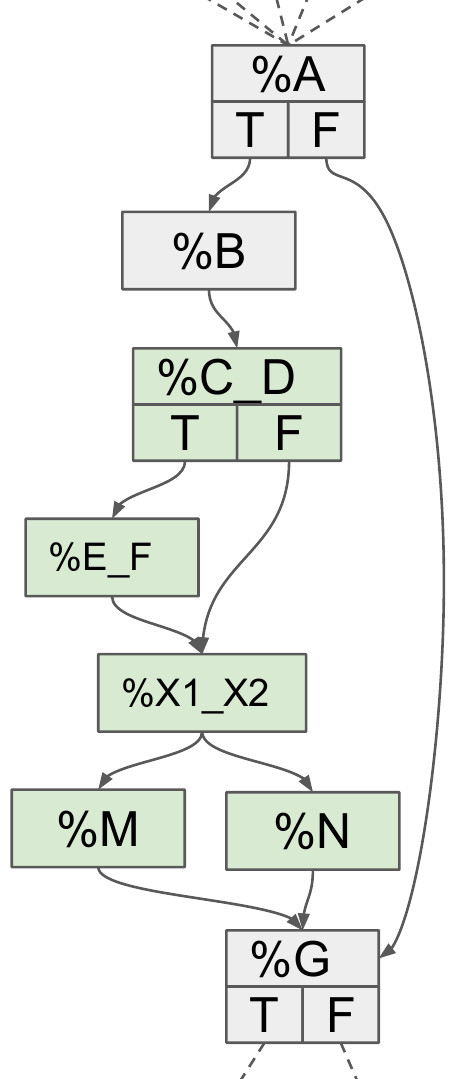}
      \caption{}
      \label{fig:melding_flow_c}
  \end{subfigure}
  \hfill
  \begin{subfigure}[b]{0.21\textwidth}
      \includegraphics[width=\textwidth]{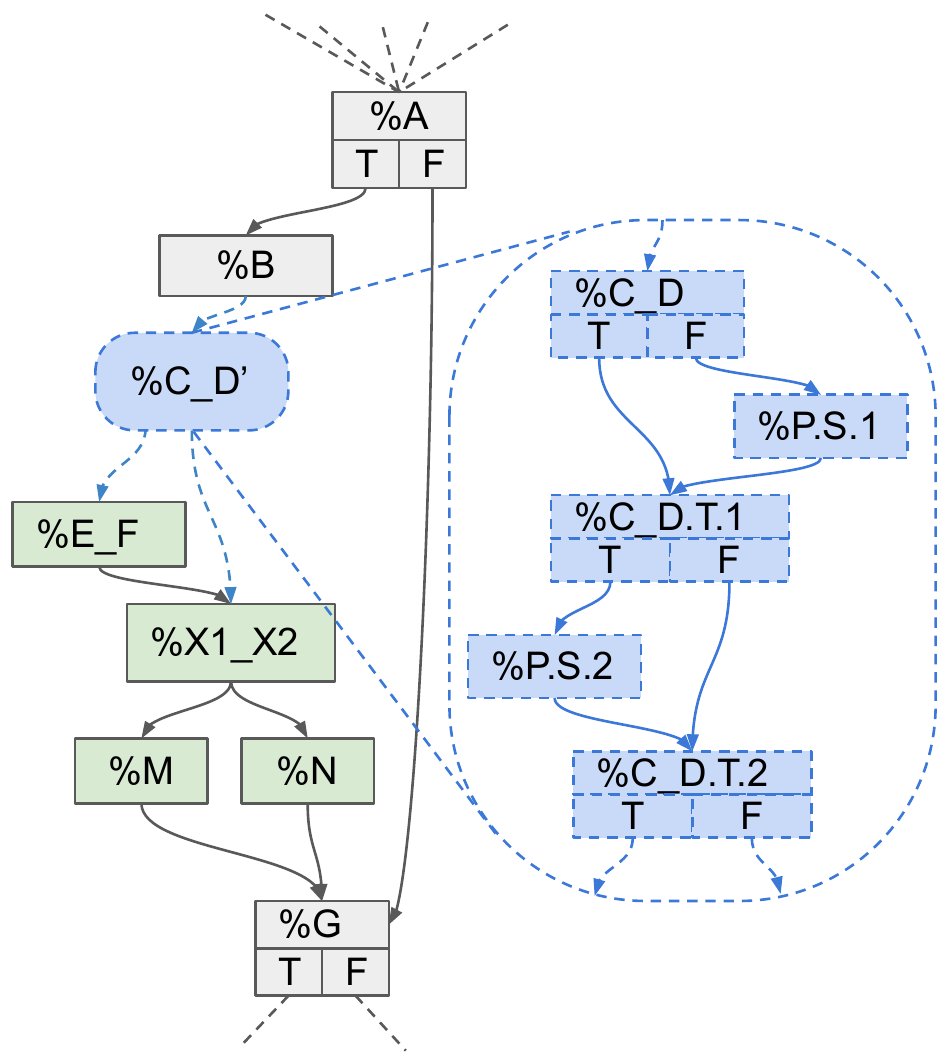}
      \caption{}
      \label{fig:melding_flow_d}
  \end{subfigure}
  \hfill
  \begin{subfigure}[b]{0.15\textwidth}
      \includegraphics[width=\textwidth]{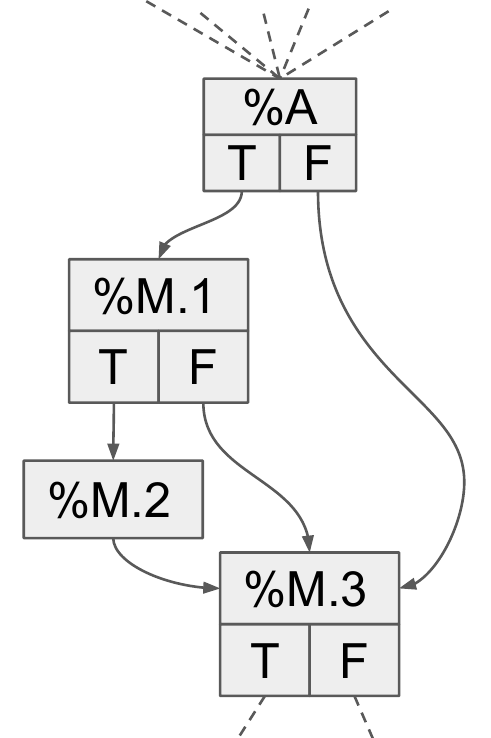}
      \caption{}
      \label{fig:melding_flow_e}
  \end{subfigure}
  \vspace{-0.5em}
  \caption{\name melding algorithm applied to bitonic sort (Figure~\ref{fig:bitonicsort_kernel})
  (a) Original control-flow graph, 
  (b) Region simplification,
  (c) \name  subgraph melding, 
  (d) Unpredication, 
  (e) Final optimized control-flow graph}
  \label{fig:melding_flow}
  \vspace{-1em}
\end{figure*}

\noindent
\textbf{Melding Branch Instructions of Exit Blocks}: Setting operands for branch instructions in subgraph  exit blocks is slightly different than that for other instructions. 
Let $B_T^{E}$,$B_F^{E}$ be the exit blocks of $S_T$ and $S_F$. Successors $B_T^{E}$,$B_F^{E}$ can contain $\phi$ nodes.
 Therefore we need to ensure that successors of $B_T^{E}$ and $B_F^{E}$ can distinguish values produced in true path or false path. To solve this we move the branch conditions of $B_T^{E}$ and $B_F^{E}$ in to newly created blocks $B_T^{'}$ and $B_F^{'}$. Now we can conditionally branch 
 to $B_T^{'}$ and $B_F^{'}$ depending on $C$. 
For example, in Figure~\ref{fig:melding_flow_c}
basic blocks $\%M$ and $\%N$ are created when when melding the exit branches of $\%X1$ and $\%X2$ in figure~\ref{fig:melding_flow_b}. Any $\phi$ node in $\%G$
(figure~\ref{fig:melding_flow_c}) can distinguish the values produced in true or false path using $\%M$ and $\%N$.

\noindent
\textbf{Melding $\phi$ Nodes} : In LLVM SSA form $\phi$ nodes are always placed at the beginning of a basic block. Even if 
the instruction alignment result contains two aligned $\phi$ nodes we can not meld them into a single $\phi$ node because  {\em select} instructions
can not be inserted before them. Therefore we copy all $\phi$ nodes into the melded basic block and set the operands for them using the $\mathit{operandMap}$.
This can introduce redundant $\phi$ nodes which we remove during post-processing.

\subsection{Unpredication}
\label{sec:design:unpred}
In our code generation process, unaligned instructions are inserted to the same melded basic block 
regardless of whether they are from true or false paths (\ie fully predicated). This can introduce overhead due to several reasons. If the branching conditions $C$ is biased 
towards the true or false path, it can result in redundant instruction execution. 
Also full predication of unaligned store instructions require adding extra loads to makes sure correct value is written back to the memory.
{\em Unpredication} splits the melded basic blocks at gap boundaries and moves the unaligned instructions into new blocks. 
Figure~\ref{fig:unpredication_example} shows unpredication applied to the unaligned instructions of basic block $B$ in Figure~\ref{fig:instr_align_example}. 
The original basic block is split to two parts ($\%M$ and $\%M.tail$) and unaligned instructions ($\%8$ and $\%9$) are moved to a new basic block, $\%M.split$.  
$\phi$ nodes (($\%10$ and $\%11$)) are added to $\%M.tail$ to ensure unaligned instructions dominate their uses.
$\%8$ and $\%9$ are never executed in the true path, therefore $\phi$ nodes' incoming values from block $\%M$ are undefined ({\em LLVM undef}). 
Note that in region replication (Section~\ref{sec:design:melding_profitability}) we apply unpredication only to the melded basic blocks. Store instructions outside the melded blocks are fully predicated by inserting extra loads.

\subsection{Pre and Post Processing Steps}
\label{sec:design:pre_and_post}
\begin{figure}[!ht]
  \vspace{-1.0em}
  \centering
  \includegraphics[width=0.35\textwidth,height=0.18\textheight]{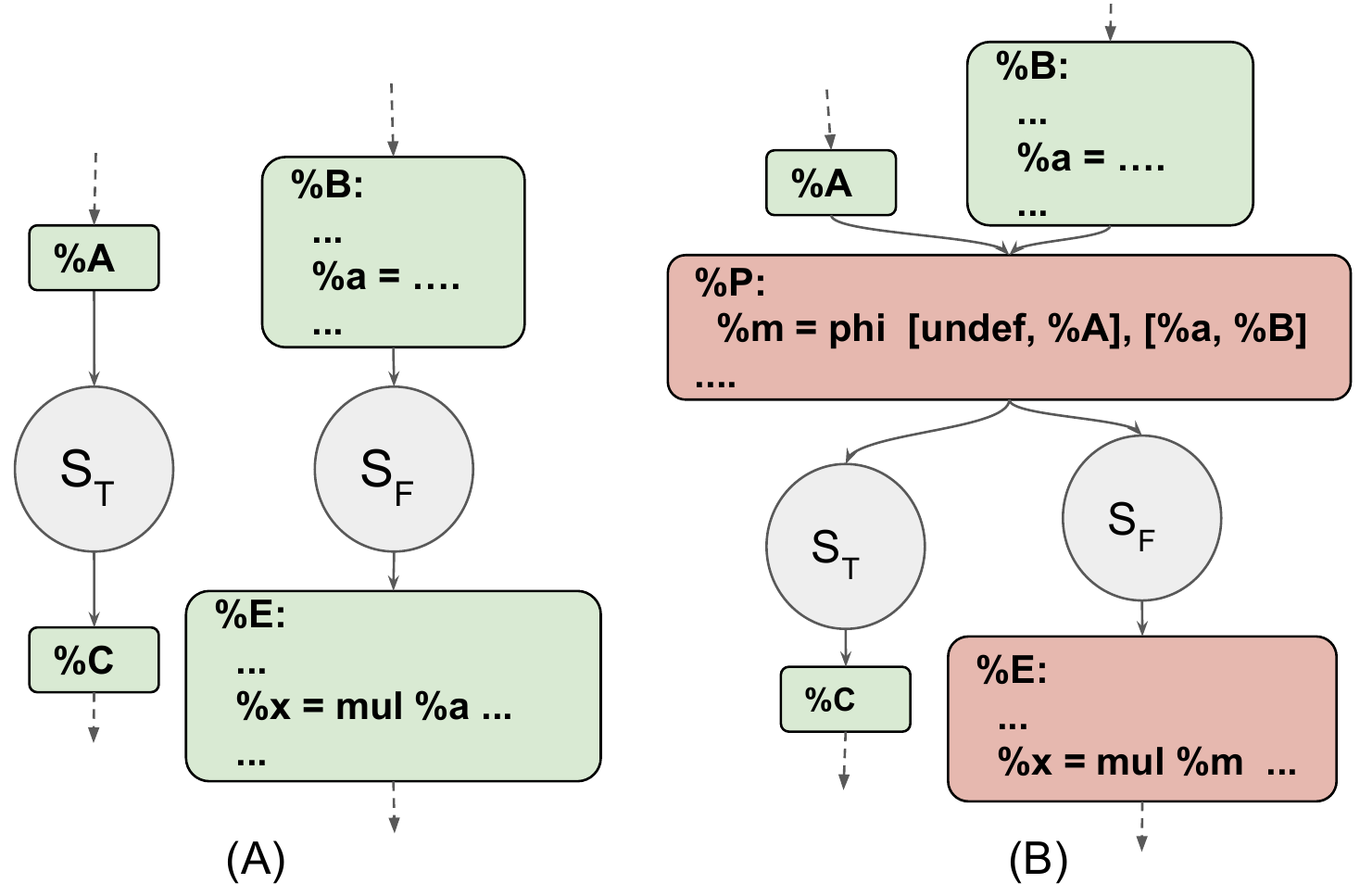}
  \vspace{-0.5em}
  \caption{\name pre-processing example}
  \label{fig:preprocess_example}
\end{figure}
In SSA form, any definition must dominate all its users. However \name's subgraph melding can break this property. Consider the two meldable subgraphs $S_T$, $S_F$  in figure~\ref{fig:preprocess_example} \circled{A}. Definition $\%a$ dominates
its use $\%x$ before the melding. However if $S_T$ and $S_F$ are melded naively then $\%a$ will no longer dominate $\%x$. To fix this we add a new basic block $\%P$ with 
a $\phi$ node $\%m$. All uses of $\%a$ are replaced with $\%m$ (Figure~\ref{fig:preprocess_example} \circled{B}). Notice that value $\%m$ is never meant to be used in the true path execution. Therefore it is 
undefined in true path ({\em undef}). We apply this preprocessing step before the melding ($\mathit{PreProcess}$ in Algorithm~\ref{algo:subgraph_melding}).

Subgraph melding can introduce branches with identical successors, $\phi$ nodes with identical operands and redundant $\phi$ nodes. 
$\mathit{RunPostOptimizations}$ in Algorithm~\ref{algo:subgraph_melding} removes these redundancies. 

\subsection{Putting All Together} Figure~\ref{fig:melding_flow} shows how each stage of the pipeline of subgraph-melding transforms the CFG of bitonicSort kernel. 
The original CFG is shown in Figure~\ref{fig:melding_flow_a}. 
Region ($\%B$, $\%G$) is a meldable divergent region. 
Figure~\ref{fig:melding_flow_b} shows the CFG after region simplification. Subgraphs $(\%C, \%X1)$ and $(\%D, \%X2)$ are profitable to meld 
according to our analysis.  
Figure~\ref{fig:melding_flow_c} shows the CFG after subgraph-melding. 
The result after applying unpredication is shown in
Figure~\ref{fig:melding_flow_d}. 
Notice that the unpredication splits the basic block $\%C\_D$ (in Figure~\ref{fig:melding_flow_c}) into 5 basic blocks (zoomed in blue-dashed blocks in Figure~\ref{fig:melding_flow_d}). 
Basic blocks $\%P.S.1$ and $\%P.S.2$ are the unaligned groups of instructions and they are executed conditionally. 
Figure~\ref{fig:melding_flow_e} shows the final optimized CFG after applying post optimizations. 
Note that {\em ROCm HIPCC} compiler applied {\em if-conversion} aggressively.
Therefore the effect of unpredication step is nullified in this case.

\begin{figure}
  \centering
    \begin{subfigure}[b]{0.3\textwidth}
        \includegraphics[width=\textwidth]{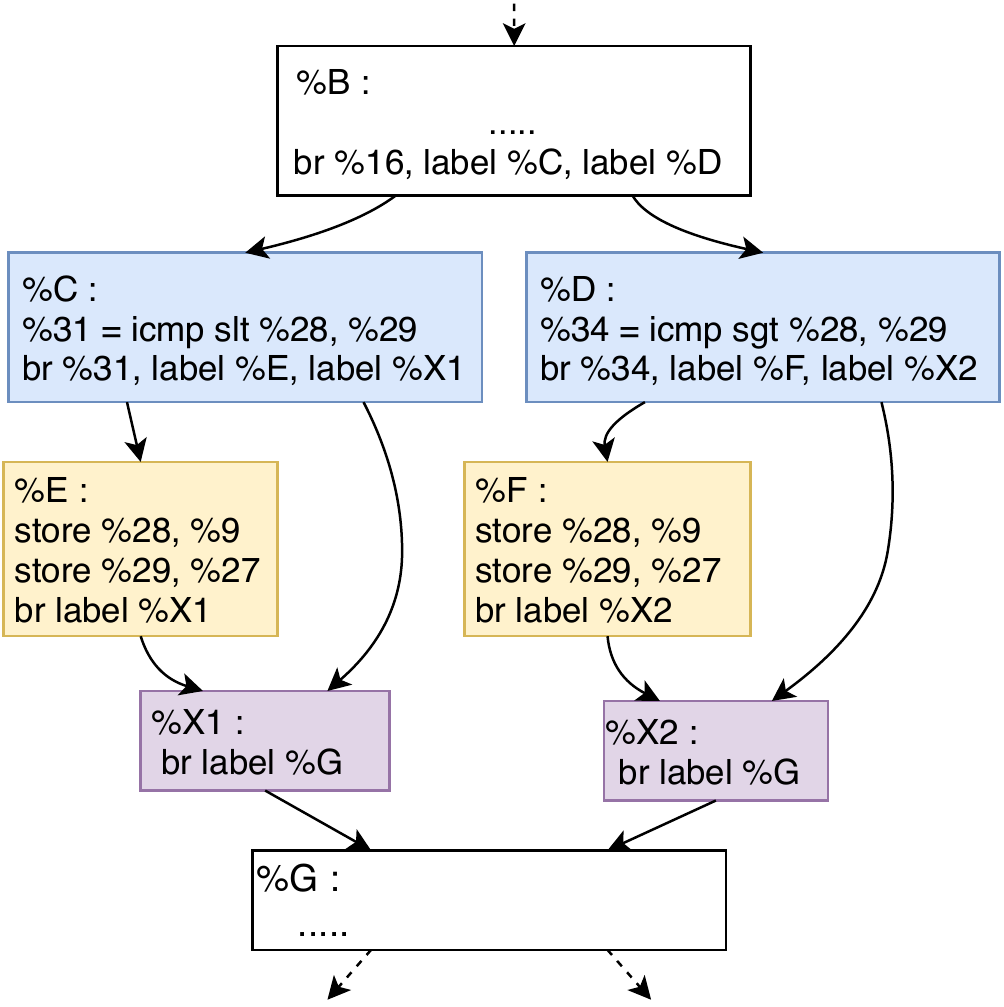}
        \caption{}
        \label{fig:llvmir_orig}
    \end{subfigure}
    \begin{subfigure}[b]{0.3\textwidth}
        \includegraphics[width=\textwidth]{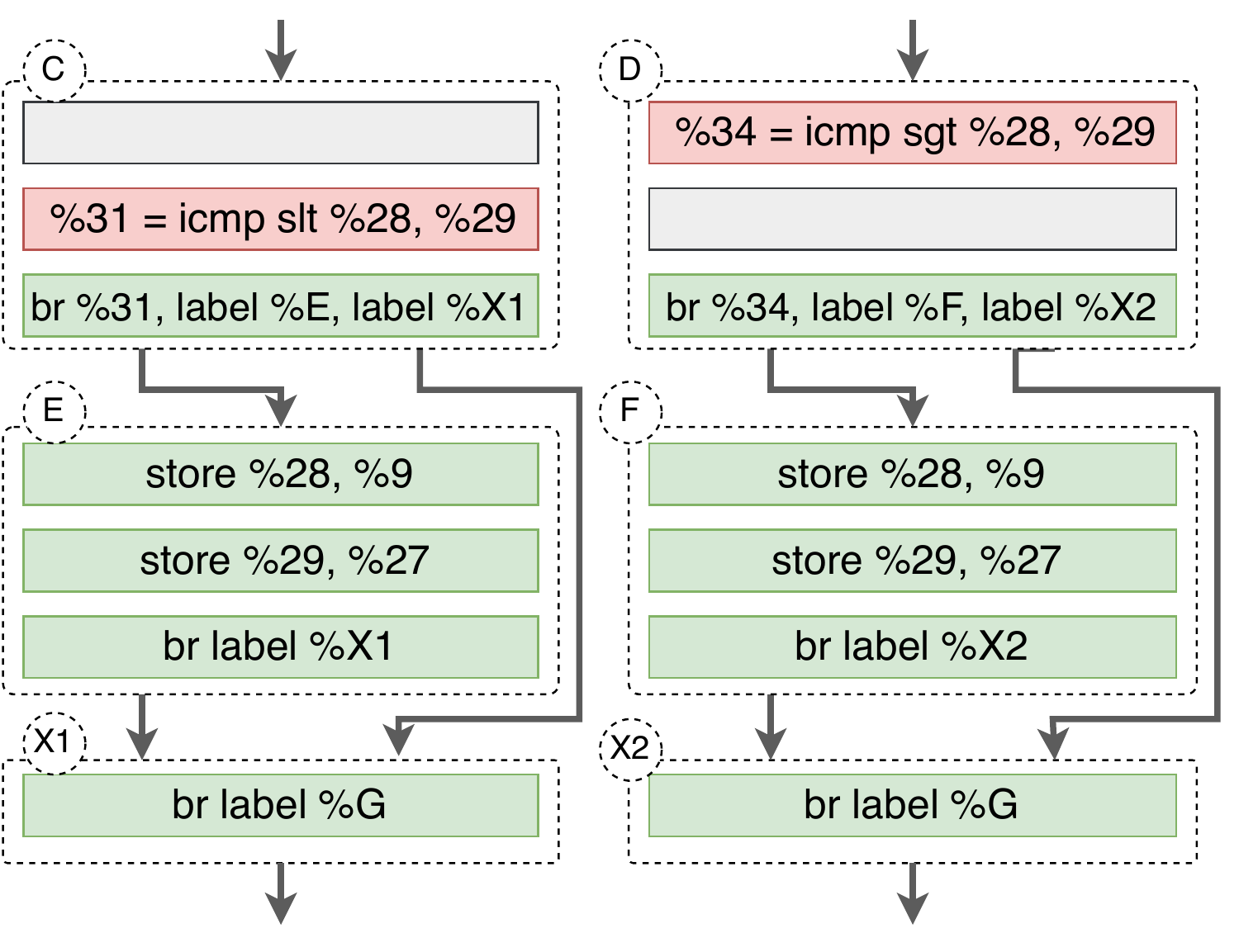}
        \caption{}
        \label{fig:llvmir_seq_align}
    \end{subfigure}
    \begin{subfigure}[b]{0.3\textwidth}
        \includegraphics[width=\textwidth]{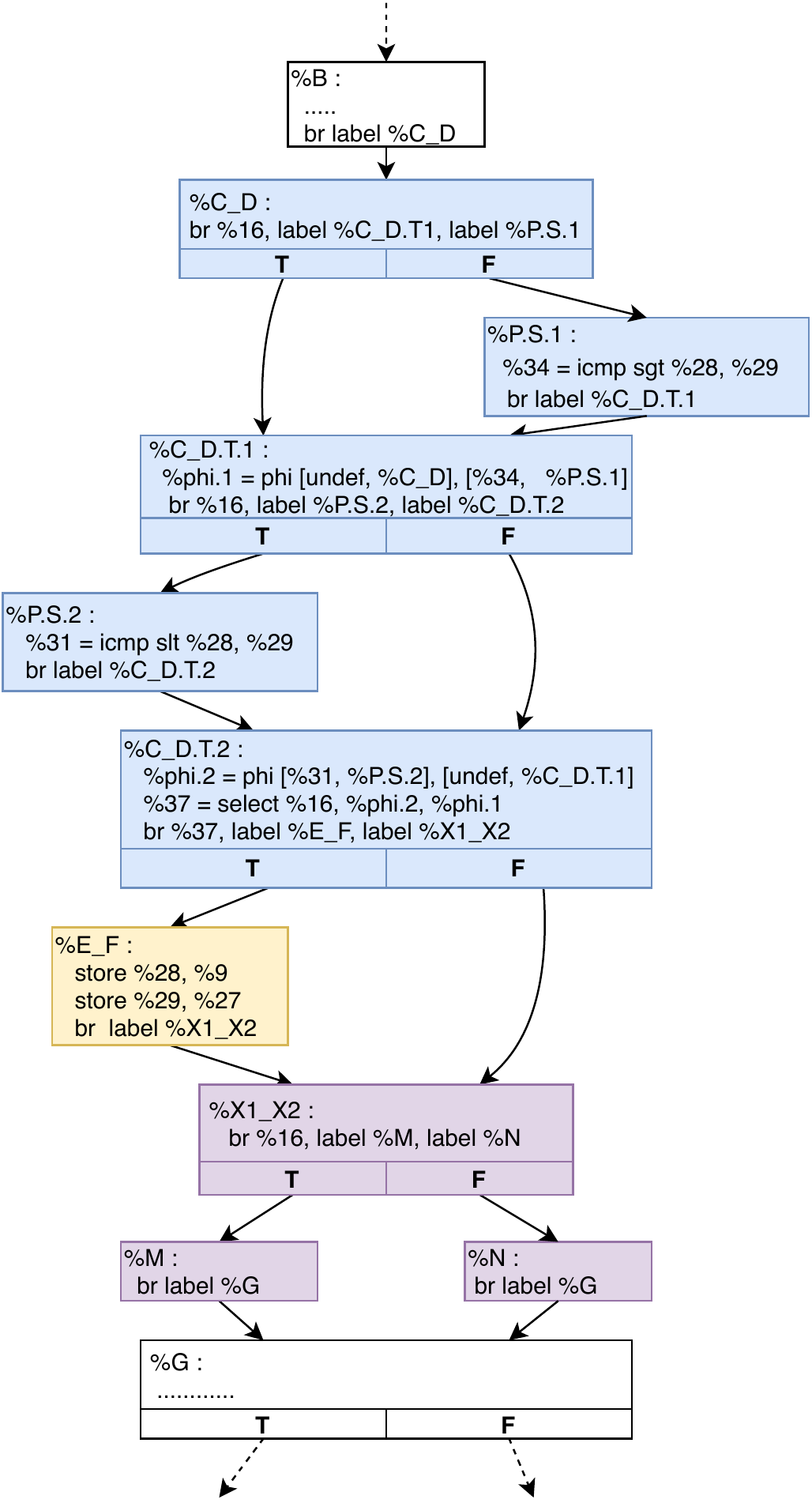}
        \caption{}
        \label{fig:llvmir_after_darm}
    \end{subfigure}
  \caption{LLVM-IR before and after applying \name transformation to our running example (a) meldable divergent region (b) instruction alignment (b) LLVM-IR generated after subgraph melding and unpredication}
  \label{fig:llvmir_example}
\end{figure}

Figure~\ref{fig:melding_flow} only shows how \name transformation changes the CFG of our running example. 
It does not show the change of instructions inside these basic blocks.
We use Figure~\ref{fig:llvmir_example} to explain the generation of melded instructions for the running example. 
Figure~\ref{fig:llvmir_orig} shows the LLVM-IR of the meldable divergent region ($(\%B,\%G)$ in Figure~\ref{fig:melding_flow_b}) in our running example.
During \name code generation, basic blocks in subgraphs $(\%C, \%X1)$ and $(\%D, \%X2)$ are linearized to compute the instruction alignment. Computed instruction alignment is shown in Figure~\ref{fig:llvmir_seq_align}. 
Notice that $[\%C,\%D],[\%E,\%F],[\%X1,\%X2]$ are the corresponding basic block pairs. 
In this example all instructions perfectly align with each other except for the compare instructions in basic blocks $\%C$ and $\%D$ (shown in red in Figure~\ref{fig:llvmir_seq_align}). 
Figure~\ref{fig:llvmir_after_darm} shows the LLVM-IR after applying subgraph melding and unpredication (similar to Figure~\ref{fig:melding_flow_d}). 
Note that instructions $\%34$ and $\%31$ (compare instructions) are unaligned. 
Therefore unpredication step introduced basic blocks $\%P.S.1$ and $\%P.S.2$ to execute them conditionally based on the divergent condition $\%16$. 
Extra $\phi$ instructions $\%phi.1$ and $\%phi.2$ are inserted to ensure def-use chains are not broken during the unpredication step. 
Out of the all aligned instructions only the branch instructions at the end of basic blocks $\%C$ and $\%D$ require select instructions during instruction-melding. 
For example the store instructions in basic blocks $\%E$, $\%F$ use matching operands, therefore can be melded without adding selects. 
On the other hand, conditional branch instructions uses values $\%34$ and $\%31$ and select instruction $\%37$ is inserted (Figure~\ref{fig:llvmir_after_darm}) to pick the branching condition {\em conditionally}. 
Note that the values $\%34$ and $\$31$ will flow to their users via the $\phi$ nodes $\%phi.1$ and $\%phi.2$ respectively. 
Therefore the select instruction (\ie $\%37$) uses these $\phi$ nodes as its operands.
 
\section{Implementation}
\label{sec:impl}
We implemented the \name algorithm described in Section~\ref{sec:design} as an LLVM-IR analysis and transformation pass
on top of the {\em ROCM HIPCC}\footnote{LLVM version 12.0.0, ROCm version 4.2.0} GPU compiler~\cite{HIPCC}. 
Both the analysis and transformation are function passes that operate on GPGPU functions. 
The analysis pass first detects meldable divergent regions using LLVM's divergence analysis. 
Then it finds all the profitable subgraph pairs that can be melded.
We use a default melding profitability threshold of 0.2 (algorithm~\ref{algo:main}). We also provide 
a sensitivity analysis on this threshold in Section~\ref{sec:melding_threshold}.
We use modified version of LLVM cost model~\cite{llvm_cost_model} to obtain instruction latencies for melding profitability
and instruction alignment computations.
The transformation uses the output of analysis to perform \name's code generation procedure (Section~\ref{sec:design:codegen}). 
The transformation pass also performs the unpredication, pre- and post-processing steps described in Sections~\ref{sec:design:unpred} and~\ref{sec:design:pre_and_post}. 
LLVM pass is implemented in $\sim 2500$ lines of C++ code.
In order to produce the program binary with our pass, we had to include our pass in the {\em ROCM HIPCC} compilation pipeline. 
Most GPGPU compilers (\eg CUDA nvcc, ROCm HIPCC) use {\em separate compilation} for GPU device and CPU host codes.
Final executable contains the device binary embedded in the host binary. 
In the modified workflow, we first compile the device code into LLVM-IR and run \name on top of that to produce a transformed IR module. 
Our pass runs only on device functions and avoids any modifications to host code. 
After that, we use the LLVM static compiler ({\em llc})~\cite{llc} to generate an object file for the transformed device code. 
The rest of the compilation flow is as same as the one without any modification.
\section{Evaluation}
\label{sec:eval}
\subsection{Evaluation Setup and Benchmarks} 
We evaluate the performance of \name on a machine with a {\em AMD Radeon Pro Vega 20} GPU. 
This GPU has 16 {\em GB}s of global memory, 64 {\em kB} of shared memory (\ie Local Data Share (LDS)) and  1700 {\em MHz} of max clock frequency. 
The machine consists of {\em AMD Ryzen Threadripper 3990X} 64-Core Processor with 2900 {\em MHz} max clock frequency. 

We use two different sets of benchmarks. 
First, to assess the generality of \name, we create several synthetic programs that exhibit control divergence of varying complexity. 
While many real-world programs are hand-optimized to eliminate divergence, these synthetic programs both qualitatively demonstrate the generality of \name over prior automated divergence-control techniques, and show that \name can automate the control flow melding that would otherwise have to be done by hand.

\noindent
\textbf{Synthetic Benchmarks} 
\begin{figure}[!ht]
  \centering
  \includegraphics[width=0.5\textwidth]{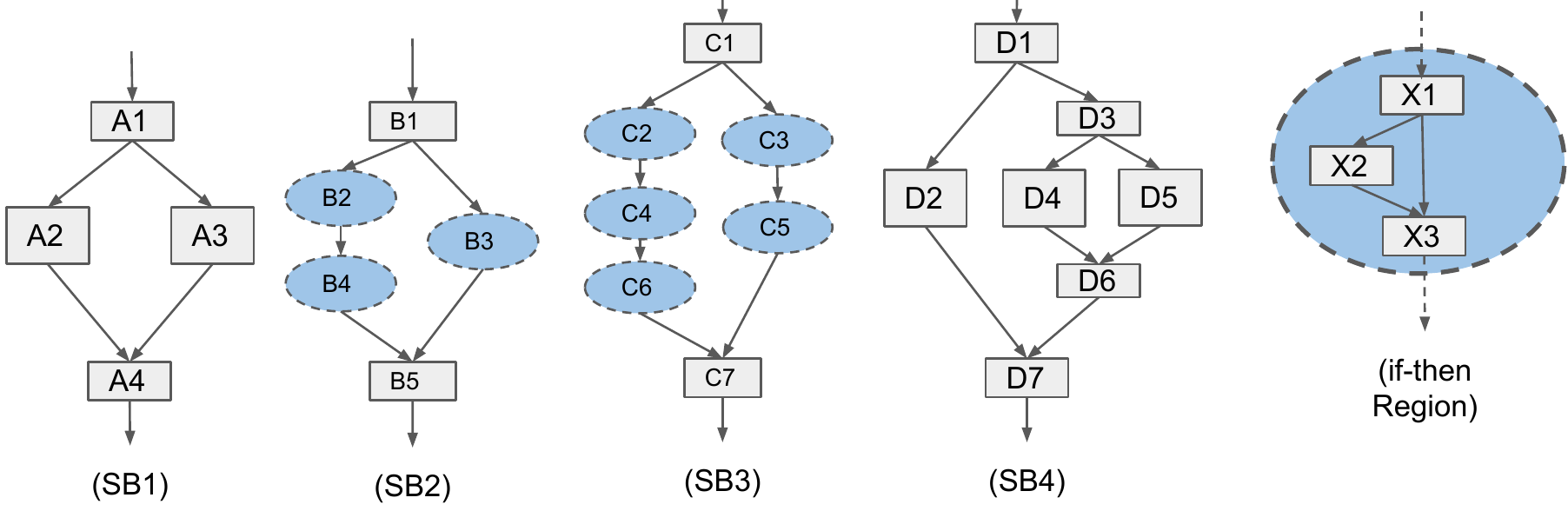}
  \caption{Control-flow patterns in synthetic benchmarks. Square: basic block and Circle: {\em if-then} region (shown on right)}
  \label{fig:sb_cases}
\end{figure}
\begin{figure*}
  \centering
  \includegraphics[width=\textwidth, height=0.18\textheight]{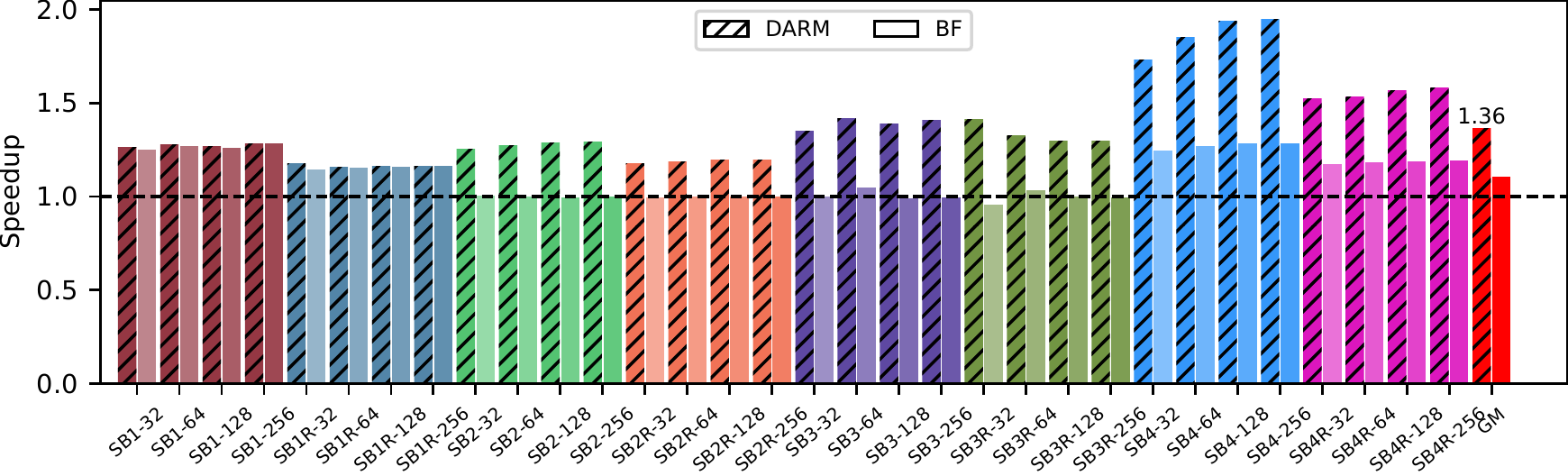}
  \caption{Micro Benchmark Performance. GM is geomean of \name's speedup over baseline.}
  \label{fig:micro_benchmark}
\end{figure*}
\begin{figure*}[!ht]
  \centering
  \includegraphics[width=\textwidth, height=0.18\textheight]{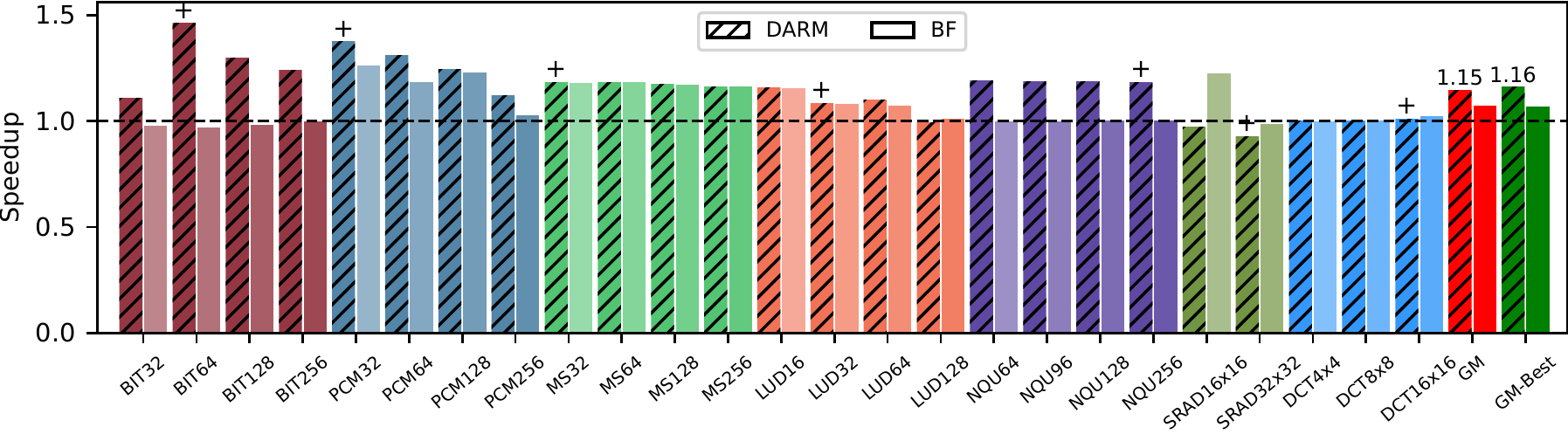}
  \vspace{-1.5em}
  \caption{Real-world Benchmark Performance. $+$ marks block size with best baseline runtime. GM is geo-mean of \name's speedup on all benchmarks; GM-Best is \name's speedup on $+$ configurations.}
  \vspace{-1.5em}
  \label{fig:real_benchmark} 
\end{figure*}
Each synthetic kernel consists of two nested loops.
The inner loop contains a divergent region with different control-flow structures (SB1, SB2, SB3 and SB4 in Figure~\ref{fig:sb_cases}).
Every divergent path computes on different pieces of data from shared memory. 
SB1 has simple diamond-shaped control-flow with basic blocks A2 and A3 performing identical computations. 
In SB2 and SB3; circled regions are {\em if-then} sections. 
{\em Then} blocks in region pairs B2-B3 (in SB2), C2-C3 and C6-C5 (in SB3) consist of identical computations.
In three-way divergent kernel SB4, basic blocks D2, D4, and D5 are performing identical computations.
Basic blocks/regions with identical computations have high melding profitability.
Synthetic benchmarks SB1-R, SB2-R, SB3-R and SB4-R have same control-flow structure as SB1-SB4 but contain non-identical computations in the 
basic blocks.
    
Prior control-flow melding techniques (tail merging~\cite{gen_tail_merge_sas03} and branch fusion~\cite{branch_fusion}) 
cannot meld the full set of synthetic benchmarks. 
Tail merging can combine the divergent {\em if-then-else} blocks in SB1 and SB4 but cannot fully merge divergent regions.
It cannot merge the -R variants due to the different instructions in the divergent paths. 
Branch fusion subsumes tail merging, and can fully merge {\em if-then-else} blocks in SB1, SB4 and their -R variants. 
However, it cannot be applied to the more complex control flow of SB2 and SB3, or their -R variants.
In SB4, iterative application of branch fusion can meld blocks D4,D5 and D2. 
However its -R variant can not be fully melded by branch fusion due to non-identical computations being un-predicated ({\em cf} Section \ref{sec:design:unpred}). 
In contrast, \name melds it by using {\em region replication} ({\em cf} Section~\ref{sec:design:melding_profitability}).

\noindent
\textbf{Real-world Benchmarks}
Second, to show \name's effectiveness on real-world programs, we consider 7 benchmarks written in {\em HIP}~\cite{HIP}. These benchmarks were taken from well-known highly hand-optimized GPU benchmark suites or optimized reference implementations of papers. We selected these benchmarks because they contain divergent if-then-else regions that present melding opportunities for DARM.  
We do not consider benchmarks that do not present any melding opportunities for \name because they are not modified by \name in any way.

  \textbf{Bitonic Sort (BIT)} Our running example is bitonic sort~\cite{bitonic}. 
  In this kernel, each thread block takes in a bucket and performs parallel sort. 
  We used an input of $2^{26}$ elements and varied the bucket (\ie block) size. 
  
  \textbf{Partition and Concurrent Merge (PCM)} PCM is a parallel sorting algorithm based on Batcher's odd-even merge sort~\cite{pcm}.
  PCM performs odd-even merging of \textit{buckets} of sorted elements at every position of the array leading to loops with nested data-dependent branches.
  We used an array of $2^{28}$ elements with different number of buckets. 
  
  \textbf{Mergesort (MS)} A parallel bottom-up merge sort implementation. 
  The kernel has data-dependent control-flow divergence in the merging step. 
  We used an input array with $2^{20}$ elements. 
  
  \textbf{LU-Decomposition (LUD)} LUD implementation from the Rodinia benchmark suite~\cite{rodinia}. 
  We focus our evaluation on the {\em lud\_perimeter} kernel in this benchmark. 
  {\em lud\_perimeter} contains multiple divergent branches that depend on thread ID and block size. 
  We use a randomly generated matrix of size $16384 \times 16384$ as the input. 
  
  \textbf{N-Queens (NQU)} N-Queens solver uses backtracking to find all different ways of placing N queens on a NxN chessboard without attacking each other. 
  We have used the kernel from the GPGPU-sim benchmark suite~\cite{ispass2009gpgpusim} with N is 15. 
  
  \textbf{Speckle Reducing Anisotropic Diffusion (SRAD)} SRAD is diffusion based noise removal method for imaging applications 
  from Rodinia benchmark suite~\cite{rodinia}. 
  We have used an image of size $4096 \times 4906$ as input.
    
  \textbf{DCT Quantization (DCT)} An in-place quantization of a discrete cosine transformation (DCT) plane~\cite{cu_samples}. 
  The quantization process is different for positive and negative values resulting in data-dependent divergence. 
  We use a randomly generated DCT plane of size $2^{15} \times 2^{15}$ as input. 
 
\noindent
\textbf{Baseline and Branch Fusion:}
Our baseline implementations of these kernels have been hand-optimized (except, obviously, for optimizations that manually remove control divergence by applying \name-like transformations). 
This optimization includes using shared memory when needed to improve performance. The baseline implementations were compiled with \nobreakdash-O3.
Branch fusion~\cite{branch_fusion} was implemented in the Ocelot~\cite{gpuocelot} open-source CUDA compiler that is no longer maintained and does not support AMD GPUs. We implemented branch fusion by modifying \name to 
apply melding for diamond-shaped control-flow ({\em if-then-else}). We use this for comparison against branch fusion.
Branch fusion cannot fully handle the control-flow of BIT, PCM, and NQU.
Loop unrolling enables successful branch fusion in LUD.

\noindent
\textbf{Block Size:}
Each of these kernels has a tunable {\em block size}---essentially, a tile size that controls the granularity of work in the inner loops. 
Because the correct block size can be dependent on many parameters (though for a given input and GPU configuration, one is likely the best), 
our evaluation treats block size as exogenous to the evaluation, and hence considers behavior at different block sizes for each kernel. 
In other words, our evaluation asks: if a programmer has a kernel with a given block size, what will happen if \name is applied?

Note that of these kernels, only LUD exhibit divergence that depends on block size. 
This means that all the other benchmarks will experience divergence regardless of block size. 
LUD's divergence, on the other hand, is block size dependent. For some block sizes, the kernel will be divergent, while for others, it will be convergent.

\subsection{Performance}
Figure~\ref{fig:micro_benchmark} shows the speedups for the synthetic benchmarks with different block sizes. 
\name can successfully meld all 4 control-flow patterns we consider in the synthetic benchmarks and gives a superior performance than the baseline and branch fusion
(geo-mean speedups of 1.36$\times$ for \name and 1.10$\times$ for branch fusion over the baseline). 
The performance for random (-R) variants are slightly lower for each of the patterns.
This is because -R variants contain random instruction sequences and instructions do not align perfectly, causing \name to insert {\em select} instructions and branches to unpredicate unaligned instruction groups. 
Speedups observed for SB3 and SB3-R are better than SB1, SB2 and their -R variants because \name melds multiple subgraph pairs in the SB3 control-flow pattern 
(Figure~\ref{fig:sb_cases}) and control-flow divergence is reduced more in this case. 
We observe the highest performance improvement for SB4 and SB4-R because \name melds basic blocks D2, D4, and D5 (Figure~\ref{fig:sb_cases}) using {\em region replication}.
SB4 and its -R variant have 3-way divergence because of the {\em if-else-if-else} branch. Applying {\em region replication} along with 
subsequent simplification passes greatly reduces this original three-way divergence. 

Figure~\ref{fig:real_benchmark} shows the speedups for real benchmarks
\name always improves the performance (1.15$\times$ geo-mean speedup over all benchmarks and 1.16$\times$ geo-mean speedup over the best baseline variants) except for SRAD (see below).
The highest relative improvement in performance can be seen in BIT and PCM for all block sizes. 
This is because both these benchmarks are divergent regardless of the block size and they have complex control-flow regions with shared memory instructions. 
\name successfully melds these regions and reduces divergence significantly. Branch fusion improves performance in 
PCM by melding {\em if-then-else} blocks.
In LUD, the divergence is block size dependent, and the kernel is divergent only at block sizes 16, 32 and 64, where we see a visible performance improvement introduced by \name.
NQU contains a time-consuming loop with divergent {\em if-then-elseif-then} section. \name applies {\em region replication} to remove divergence, 
achieving superior performance.
SRAD kernel has both block size-dependent and data-dependent divergent regions (say $R_B$ and $R_D$ respectively).
Both $R_B$ and $R_D$ consists of {\em if-then-else--if-then-else} chains.
$R_B$ contains no shared memory instructions and melding does not improve performance (for both \name and branch fusion). 
However $R_D$ contains a 3-way divergent branch with shared memory instructions and the divergence is biased \ie 
execution only takes 2 of the 3 ways. 
In this case branch fusion has better performance at block size 16, because blocks that get melded happen to be on the divergent paths. 
However \name has more melding options than branch fusion, and it melds all 3 paths adding extra overhead. 
At block size 32, the extra overhead introduced by melding $R_B$ becomes significant and both \name and branch fusion exhibit a performance drop.
Performance drop for \name can be avoided by prioritizing the melding order (\ie apply melding to divergent regions with most profitable subgraphs first). 
However, prioritizing melding order is not considered in this paper.

In most cases (except SRAD), the block size for best performing baseline is also the one that gives the best absolute performance for \name.
Interestingly, for $4/7$ benchmarks (BIT, PCM, MS, and DCT), not only does this best baseline block size produce the best absolute \name performance, it also produces the best {\em speedup} relative to the baseline: the block size that makes the baseline perform the best, actually exposes more optimization opportunities to \name. 

We  use {\em rocprof}~\cite{rocprof} to collect ALU utilization and memory instruction counters to reason about performance.
We focus on the block sizes for each benchmark where \name has highest improvement over the baseline.

\subsection{ALU Utilization}
\begin{figure}[!b]
  \centering
  \includegraphics[width=0.45\textwidth, height=0.13\textheight]{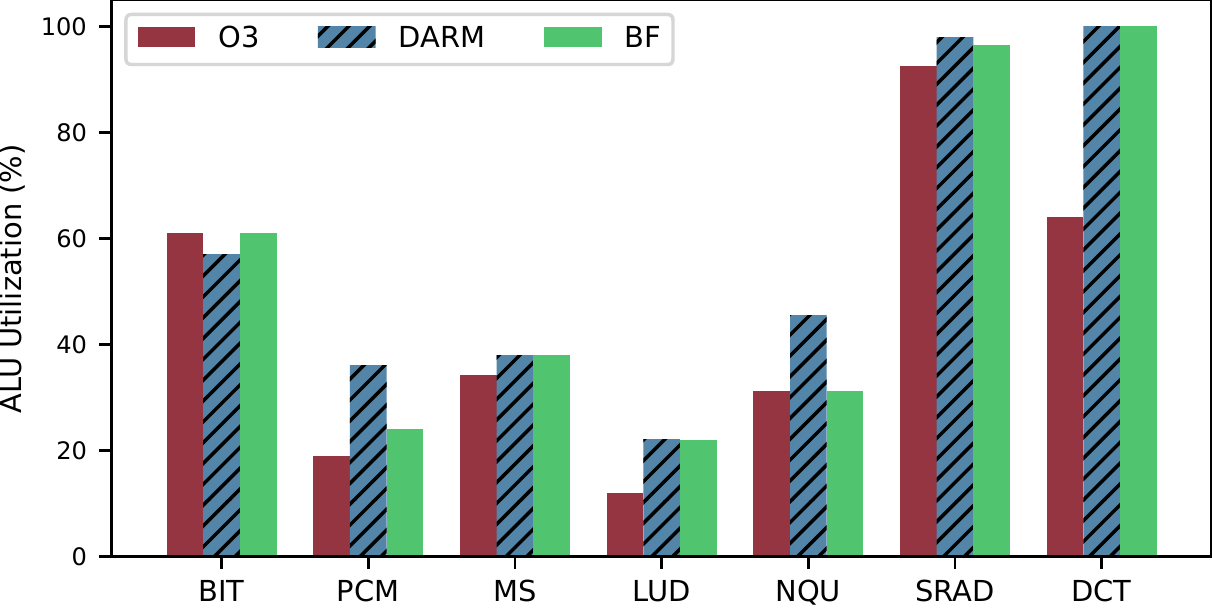}
  \caption{ALU Utilization.}
  \label{fig:alu_util}
\end{figure}

\name's melding transformation enables the ALU instructions in divergent paths to be issued in the same cycle. 
This effectively improves the SIMD resource utilization. 
Figure~\ref{fig:alu_util} shows the ALU utilization ($\%$).
As expected \name improves the ALU utilization significantly for most benchmarks. 
In BIT, divergent paths does not have common comparison operators ($>$ and $<$ comparisons in lines 9 and 13 in Figure~\ref{fig:bitonicsort_kernel}). 
Even though \name unpredicates these instructions, later optimization passes decide to fully-predicate them resulting in lower ALU utilization.
\begin{figure}[!b]
  \centering
  \includegraphics[width=0.45\textwidth, height=0.13\textheight]{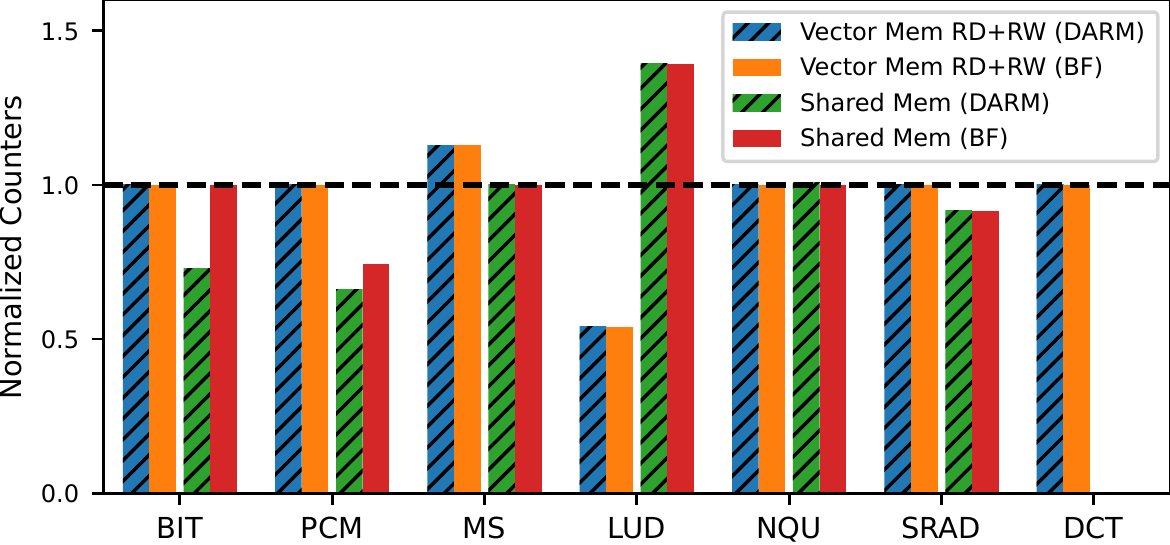}
  \caption{Normalized Memory Instruction Counters.}
  \label{fig:mem_util}
\end{figure}

\subsection{Melding of Memory Instructions}
\label{sec:eval:memory}

Figure~\ref{fig:mem_util} shows the normalized number of 
global and shared memory (\ie local data share) instructions issued after applying \name. 
In LUD, there are many common shared memory instructions in divergent paths. 
However these instructions do not have different memory alignments, therefore cannot be melded into a single instruction. 
Unpredicated shared memory instructions are predicted by other optimization passes in LLVM resulting in higher instruction count.
Melding reduces the global memory instruction count in LUD.
DCT does not have any memory instructions in the divergent region and does not use shared memory.
In BIT and PCM, the melded regions contain a lot of shared memory instructions. Therefore the reduction in shared memory instructions is significant and correlate with the performance gain. 
We find that melding shared memory instructions is more beneficial than melding ALU instructions because shared memory instructions have higher latency than most ALU instructions, though lower latency than global memory instructions. 
Therefore there is 2$\times$ improvement in cycles spent if 
two divergent shared memory instructions are issued in the same cycle. 
In contrast, melding global memory instructions does not always improve performance. 
This is because the data requested by divergent memory instructions might be on different cache lines and 
these requests are serialized by the memory controller even if they are issued in the same cycle.
\subsection{Melding Profitability Threshold}
\label{sec:melding_threshold}
\begin{figure}[!t]
  \centering
  \includegraphics[width=0.45\textwidth, height=0.13\textheight]{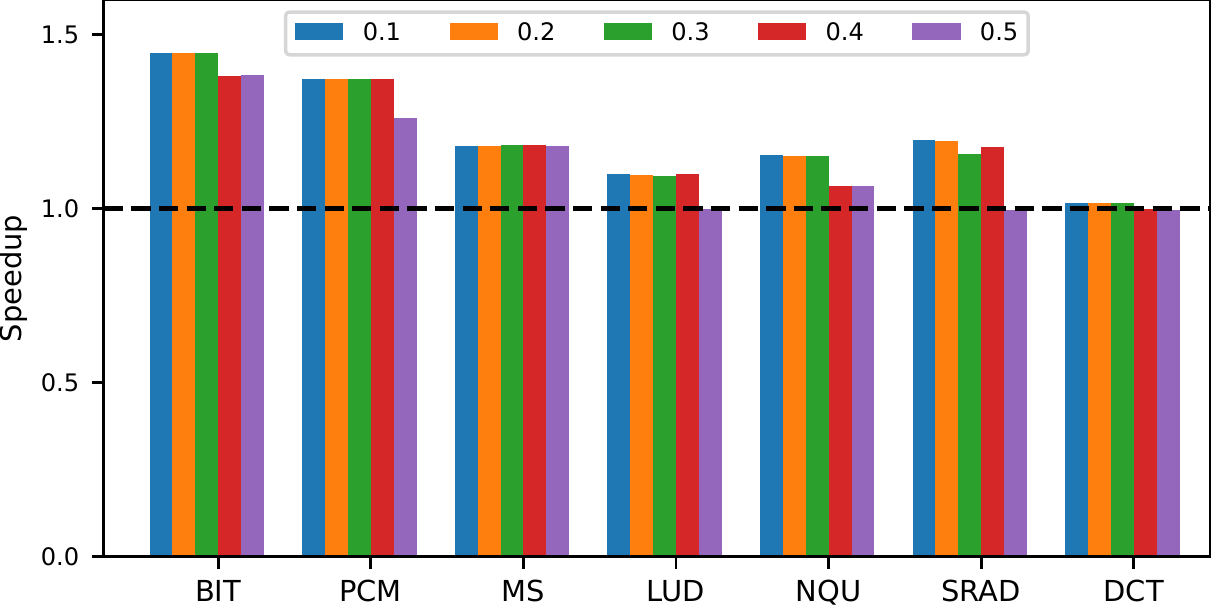}
  \caption{Variation of melding profitability thresholds.}
  \label{fig:profit}
\end{figure}

Figure~\ref{fig:profit} shows the performance of \name for different melding profitability thresholds on the real-world benchmarks considering \name's best performing block sizes.
For all benchmarks, we observe that \name's speedup reduces as we increase the threshold due to lost opportunities.
When we reduce the threshold, increment in the improvement of the performance of \name becomes insignificant (after 0.2).
But we cannot reduce it to zero because every possible pair would be melded and the subsequent CFG simplification passes would unpredicate them.
As a result, \name may become non-convergent. 

\subsection{Compile Time}
\begin{table}[!ht]
  \centering
  \caption{Average Compile Time (s)}
  \label{tab:compile-time}
  \small
  \resizebox{0.25\textwidth}{!}{
  \begin{tabular}{lrrr}\toprule
    Benchmark &O3 &DARM &Normalized \\\midrule
    BIT  &0.4804 &0.5018 &1.0444 \\
    PCM  &0.5690 &0.5942 &1.0443 \\
    MS   &0.8037 &0.8064 &1.0035 \\
    LUD  &0.5993 &0.6294 &1.0502 \\
    NQU  &0.4687 &0.4738 &1.0109 \\
    SRAD &0.4999 &0.5121 &1.0244 \\
    DCT  &0.4398 &0.4439 &1.0093 \\
    \bottomrule
  \end{tabular}}
\end{table}
Table~\ref{tab:compile-time} shows the device code compilation times for the baseline and \name. 
We omit the time for compiling host code and linking because it is constant for 
both the baseline and \name. 
Since we perform the analysis and the instruction alignment -- the most costly parts -- at the basic block level rather than performing at a higher level (\ie function or region level), we incur negligible compilation overhead. Compilation time overhead introduced by \name is a small fraction of total compilation time (including host code) for all cases.

\name's compile time depends on the size of basic blocks that get melded and the structure of the program since it determines different types of melding opportunities. 
A slight overhead in compilation time of LUD is caused by sequence alignment overhead on large basic blocks (created by loop unrolling).
PCM and BIT have divergent regions inside an unrolled loop, therefore \name's meldable subgraph detection incurs overhead.
Only BIT and PCM has opportunities for Region-Region melding, and only PCM, NQU, and SRAD have opportunities for Basic block-Region melding.
Presence of Basic block-Region melding opportunity results in {\em region replication}.
\section{Related Work}
\label{sec:related_work}


\paragraph{Divergence Analysis}
Impact of control-flow divergence has extensively studied in different contexts~\cite{taco2015hack,wholefuncv2011,pldi2018hack,gpucheck}.
Reducing control-flow divergence requires finding the source of divergence in a program.
Coutinho~\etal constructed a divergence analysis to statically identify variables with the same value for every SIMD unit and used this analysis to drive {\em Branch Fusion}~\cite{branch_fusion}.
A divergence analysis of similar fashion based on data and sync dependences has been integrated to the LLVM framework~\cite{llvm_div_analysis}.
Recently, Rosemann~\etal has presented a precise divergence analysis based on abstract interpretation for reducible CFGs~\cite{divanalysispopl2021}.
Using a precise divergence analysis improves the opportunities of melding for \name.

\paragraph{Code Compaction}
\emph{Tail Merging} is a standard, but restrictive, compiler optimization used to reduce the code size by merging identical sequences of instructions. 
Chen~\etal used generalized tail merging to compact matching \emph{Single-Entry-Multiple-Exit} regions~\cite{gen_tail_merge_sas03}.
Recently, Rocha~\etal has presented \emph{Function Merging}, an advanced sequence-alignment based technique for code size reduction~\cite{rocha20, rocha19}.
Even though parts of \name has some similarities with function merging, it does not tackle divergence.

\paragraph{Compiler Techniques}
In addition to branch fusion, Anantpur and Govindarajan proposed to structure the unstructured CFGs and then linearize it with predication~\cite{linearcfg2014}.
More recently, Fukuhara and Takimoto proposed \emph{Speculative Sparse Code Motion} to reduce divergence in GPU programs~\cite{codemotion20},
which preserves the CFG and it is orthogonal to \name.
\emph{Collaborative Context Collection} copies registers of divergent warps to shared memory and restores them when those warps become non-divergent~\cite{ccc2015}.
\emph{Iteration Delaying} is a complementary compiler optimization to \name that delays divergent loop iterations~\cite{iterdelay2011} and can be applied following \name.
Recently, Damani~\etal has presented a speculative reconvergence technique for GPUs similar to iteration delaying~\cite{spec_reconv_cgo20}.
\emph{Common Subexpression Convergence (CSC)}~\cite{common_subexpression_convergence} works similar to branch fusion but uses {\em branch flattening} (\ie predication) to handle complex control-flow.
In contrast, \name does not require predication to meld complex control-flow, thus more general than CSC.

\paragraph{Architectural Techniques}
Thread Block Compaction~\cite{tblockcompact} and Dynamic Warp Formation~\cite{dynamicwarp} involve repacking threads into non-divergent warps.
Variable Warp Sizing~\cite{varwarp} and Dynamic Warp Subdivision~\cite{warpsubdivision} depend on smaller warps to schedule divergent thread groups in parallel.
Independent Thread Scheduling helps to hide the latency in divergent paths by allowing to switch between divergent threads inside a warp~\cite{dualpath,multipath}.
\section{Discussion and Future Work}
\label{sec:discussion}
Most of the GPGPU benchmarks are heavily hand optimized by expert developers and this often include \name like transformations to remove control-flow divergence~\cite{branch_fusion}. We evaluate \name on limited set of real-world benchmarks mainly because of this reason. However we also emphasize that doing \name-like transformations by hand is time-consuming and error-prone. 
For example, it took us several hours to manually apply control-flow melding to LUD kernel.  
Therefore, offloading this to the compiler can save a lot of developer effort.

The benefits of \name is not limited to reducing control-flow divergence in GPGPU programs. \name can be used to reduce control-flow divergence in any hardware backends and programming models that employ SIMT execution (\eg intel/AMD processors with ISPC~\cite{ispc}). \name can be used to reduce branches in a program. This property can be exploited to accelerate software testing techniques such as symbolic execution~\cite{targeted_transformations}.
\name factor out common code segments within {\em if-the-else} regions of a program.
Therefore it can be used as an intra-function code size reduction optimization as well.
Aforementioned applications of \name suggest that it is useful as a general compiler optimization technique. We plan to explore some of these applications in our future work.  


In Section~\ref{sec:eval}, we have shown that when shared memory is used to improve the baseline, it does not steal the opportunity from \name to meld, because melding shared memory instructions also results in better performance than the improved baseline.
Exploiting this opportunity requires maximizing the alignment of shared memory instructions which can be achieved by using a refined instruction cost model.

\section{Conclusion}
\label{sec:conclusion}
Divergent control-flow in GPGPU programs causes performance degradation due to serialization. 
We presented \name, a new compiler analysis and transformation framework for GPGPU programs implemented on LLVM, that can detect and meld similar control-flow regions in divergent paths to reduce divergence in control-flow. 
\name generalizes and subsumes prior efforts at reducing divergence such as tail merging and branch fusion.
We showed that \name improves performance by improving ALU utilization and promoting coalesced shared memory accesses across several real-world benchmarks.
\section*{Acknowledgments}
This work was supported in part by National Science Foundation awards CCF-1919197 and CCF-1908504. 
We  would like to thank anonymous reviewers for their helpful comments and feedback.
We would like to thank Tim Rogers for his feedback during discussions of this work and also providing us AMD GPUs for the experiments. Furthermore, we would like to thank Rodrigo Rocha for sharing the source code for Function Merging.

\bibliographystyle{IEEEtran}
\balance
\bibliography{references}

\begin{thebibliography}{10}
\providecommand{\url}[1]{#1}
\csname url@samestyle\endcsname
\providecommand{\newblock}{\relax}
\providecommand{\bibinfo}[2]{#2}
\providecommand{\BIBentrySTDinterwordspacing}{\spaceskip=0pt\relax}
\providecommand{\BIBentryALTinterwordstretchfactor}{4}
\providecommand{\BIBentryALTinterwordspacing}{\spaceskip=\fontdimen2\font plus
\BIBentryALTinterwordstretchfactor\fontdimen3\font minus
  \fontdimen4\font\relax}
\providecommand{\BIBforeignlanguage}[2]{{%
\expandafter\ifx\csname l@#1\endcsname\relax
\typeout{** WARNING: IEEEtran.bst: No hyphenation pattern has been}%
\typeout{** loaded for the language `#1'. Using the pattern for}%
\typeout{** the default language instead.}%
\else
\language=\csname l@#1\endcsname
\fi
#2}}
\providecommand{\BIBdecl}{\relax}
\BIBdecl

\bibitem{dynamicwarp}
W.~W.~L. {Fung}, I.~{Sham}, G.~{Yuan}, and T.~M. {Aamodt}, ``Dynamic warp
  formation and scheduling for efficient gpu control flow,'' in \emph{40th
  Annual IEEE/ACM International Symposium on Microarchitecture (MICRO 2007)},
  2007, pp. 407--420.

\bibitem{compact_thblk}
W.~W.~L. {Fung} and T.~M. {Aamodt}, ``Thread block compaction for efficient
  simt control flow,'' in \emph{2011 IEEE 17th International Symposium on High
  Performance Computer Architecture}, 2011, pp. 25--36.

\bibitem{dualpath}
M.~{Rhu} and M.~{Erez}, ``The dual-path execution model for efficient gpu
  control flow,'' in \emph{2013 IEEE 19th International Symposium on High
  Performance Computer Architecture (HPCA)}, 2013, pp. 591--602.

\bibitem{gen_tail_merge_sas03}
W.-K. Chen, B.~Li, and R.~Gupta, ``Code compaction of matching single-entry
  multiple-exit regions,'' in \emph{Proceedings of the 10th International
  Conference on Static Analysis}, ser. SAS'03.\hskip 1em plus 0.5em minus
  0.4em\relax Berlin, Heidelberg: Springer-Verlag, 2003, p. 401–417.

\bibitem{branch_fusion}
B.~{Coutinho}, D.~{Sampaio}, F.~M.~Q. {Pereira}, and W.~{Meira Jr.},
  ``Divergence analysis and optimizations,'' in \emph{2011 International
  Conference on Parallel Architectures and Compilation Techniques}, 2011, pp.
  320--329.

\bibitem{llvm}
C.~{Lattner} and V.~{Adve}, ``Llvm: a compilation framework for lifelong
  program analysis transformation,'' in \emph{International Symposium on Code
  Generation and Optimization, 2004. CGO 2004.}, 2004, pp. 75--86.

\bibitem{HIP}
\BIBentryALTinterwordspacing
``{HIP Programming Guide v4.1},'' [Accessed 17-Dec-2021]. [Online]. Available:
  \url{https://rocmdocs.amd.com/en/latest/}
\BIBentrySTDinterwordspacing

\bibitem{cuda}
\BIBentryALTinterwordspacing
``{CUDA C++ Programming Guide},'' [Accessed 17-Dec-2021]. [Online]. Available:
  \url{https://docs.nvidia.com/cuda/cuda-c-programming-guide/index.html}
\BIBentrySTDinterwordspacing

\bibitem{nvcc}
\BIBentryALTinterwordspacing
``{NVCC :: CUDA Toolkit Documentation},'' [Accessed 17-Dec-2021]. [Online].
  Available:
  \url{https://docs.nvidia.com/cuda/cuda-compiler-driver-nvcc/index.html}
\BIBentrySTDinterwordspacing

\bibitem{HIPCC}
\BIBentryALTinterwordspacing
``{ROCm Compiler SDK},'' [Accessed 17-Dec-2021]. [Online]. Available:
  \url{https://rocmdocs.amd.com/en/latest/ROCm_Compiler_SDK/ROCm-Compiler-SDK.html}
\BIBentrySTDinterwordspacing

\bibitem{ssa_form}
\BIBentryALTinterwordspacing
R.~Cytron, J.~Ferrante, B.~K. Rosen, M.~N. Wegman, and F.~K. Zadeck,
  ``Efficiently computing static single assignment form and the control
  dependence graph,'' \emph{ACM Trans. Program. Lang. Syst.}, vol.~13, no.~4,
  p. 451–490, Oct. 1991. [Online]. Available:
  \url{https://doi.org/10.1145/115372.115320}
\BIBentrySTDinterwordspacing

\bibitem{llvm_div_analysis}
R.~Karrenberg and S.~Hack, ``Improving performance of opencl on cpus,'' in
  \emph{Compiler Construction}, M.~O'Boyle, Ed.\hskip 1em plus 0.5em minus
  0.4em\relax Berlin, Heidelberg: Springer Berlin Heidelberg, 2012, pp. 1--20.

\bibitem{divanalysispopl2021}
\BIBentryALTinterwordspacing
J.~Rosemann, S.~Moll, and S.~Hack, ``An abstract interpretation for spmd
  divergence on reducible control flow graphs,'' \emph{Proc. ACM Program.
  Lang.}, vol.~5, no. POPL, Jan. 2021. [Online]. Available:
  \url{https://doi.org/10.1145/3434312}
\BIBentrySTDinterwordspacing

\bibitem{bitonic}
K.~E. {Batcher}, ``Sorting networks and their applications,'' in
  \emph{Proceedings of the April 30--May 2, 1968, spring joint computer
  conference (AFIPS '68 (Spring))}, 1968, p. 307–314.

\bibitem{gpu_quicksort_cederman}
\BIBentryALTinterwordspacing
D.~Cederman and P.~Tsigas, ``Gpu-quicksort: A practical quicksort algorithm for
  graphics processors,'' \emph{ACM J. Exp. Algorithmics}, vol.~14, Jan. 2010.
  [Online]. Available: \url{https://doi.org/10.1145/1498698.1564500}
\BIBentrySTDinterwordspacing

\bibitem{llvm_region}
\BIBentryALTinterwordspacing
``{llvm::RegionBase Class Template Reference},'' [Accessed 17-Dec-2021].
  [Online]. Available:
  \url{https://llvm.org/doxygen/classllvm_1_1RegionBase.html}
\BIBentrySTDinterwordspacing

\bibitem{structure_tree}
\BIBentryALTinterwordspacing
R.~Johnson, D.~Pearson, and K.~Pingali, ``The program structure tree: Computing
  control regions in linear time,'' \emph{SIGPLAN Not.}, vol.~29, no.~6, p.
  171–185, Jun. 1994. [Online]. Available:
  \url{https://doi.org/10.1145/773473.178258}
\BIBentrySTDinterwordspacing

\bibitem{warp_primitives}
\BIBentryALTinterwordspacing
``Using cuda warp-level primitives,'' [Accessed 17-Dec-2021]. [Online].
  Available:
  \url{https://developer.nvidia.com/blog/using-cuda-warp-level-primitives/}
\BIBentrySTDinterwordspacing

\bibitem{smith_waterman}
\BIBentryALTinterwordspacing
T.~Smith and M.~Waterman, ``Identification of common molecular subsequences,''
  \emph{Journal of Molecular Biology}, vol. 147, no.~1, pp. 195--197, 1981.
  [Online]. Available:
  \url{https://www.sciencedirect.com/science/article/pii/0022283681900875}
\BIBentrySTDinterwordspacing

\bibitem{rocha19}
R.~C.~O. {Rocha}, P.~{Petoumenos}, Z.~{Wang}, M.~{Cole}, and H.~{Leather},
  ``Function merging by sequence alignment,'' in \emph{2019 IEEE/ACM
  International Symposium on Code Generation and Optimization (CGO)}, 2019, pp.
  149--163.

\bibitem{rocha20}
\BIBentryALTinterwordspacing
R.~C.~O. Rocha, P.~Petoumenos, Z.~Wang, M.~Cole, and H.~Leather, ``Effective
  function merging in the ssa form,'' in \emph{Proceedings of the 41st ACM
  SIGPLAN Conference on Programming Language Design and Implementation}, ser.
  PLDI 2020.\hskip 1em plus 0.5em minus 0.4em\relax New York, NY, USA:
  Association for Computing Machinery, 2020, p. 854–868. [Online]. Available:
  \url{https://doi.org/10.1145/3385412.3386030}
\BIBentrySTDinterwordspacing

\bibitem{llvm_cost_model}
\BIBentryALTinterwordspacing
``{CostModel.cpp File Reference},'' [Accessed 17-Dec-2021]. [Online].
  Available: \url{https://llvm.org/doxygen/CostModel_8cpp.html}
\BIBentrySTDinterwordspacing

\bibitem{llc}
\BIBentryALTinterwordspacing
``{llc - LLVM static compiler},'' [Accessed 17-Dec-2021]. [Online]. Available:
  \url{https://llvm.org/docs/CommandGuide/llc.html}
\BIBentrySTDinterwordspacing

\bibitem{pcm}
E.~{Herruzo}, G.~{Ruiz}, J.~I. {Benavides}, and O.~{Plata}, ``A new parallel
  sorting algorithm based on odd-even mergesort,'' in \emph{15th EUROMICRO
  International Conference on Parallel, Distributed and Network-Based
  Processing (PDP'07)}, 2007, pp. 18--22.

\bibitem{rodinia}
S.~{Che}, M.~{Boyer}, J.~{Meng}, D.~{Tarjan}, J.~W. {Sheaffer}, S.~{Lee}, and
  K.~{Skadron}, ``Rodinia: A benchmark suite for heterogeneous computing,'' in
  \emph{2009 IEEE International Symposium on Workload Characterization
  (IISWC)}, 2009, pp. 44--54.

\bibitem{ispass2009gpgpusim}
A.~Bakhoda, G.~L. Yuan, W.~W.~L. Fung, H.~Wong, and T.~M. Aamodt, ``Analyzing
  cuda workloads using a detailed gpu simulator,'' in \emph{2009 IEEE
  International Symposium on Performance Analysis of Systems and Software},
  2009, pp. 163--174.

\bibitem{cu_samples}
\BIBentryALTinterwordspacing
``{CUDA Samples},'' [Accessed 17-Dec-2021]. [Online]. Available:
  \url{https://docs.nvidia.com/cuda/cuda-samples/}
\BIBentrySTDinterwordspacing

\bibitem{gpuocelot}
A.~Kerr, G.~Diamos, and S.~Yalamanchili, ``A characterization and analysis of
  ptx kernels,'' in \emph{2009 IEEE International Symposium on Workload
  Characterization (IISWC)}, 2009, pp. 3--12.

\bibitem{rocprof}
\BIBentryALTinterwordspacing
``{ ROCm-Developer-Tools / rocprofiler },'' [Accessed 17-Dec-2021]. [Online].
  Available: \url{https://github.com/ROCm-Developer-Tools/rocprofiler}
\BIBentrySTDinterwordspacing

\bibitem{taco2015hack}
\BIBentryALTinterwordspacing
T.~Schaub, S.~Moll, R.~Karrenberg, and S.~Hack, ``The impact of the simd width
  on control-flow and memory divergence,'' \emph{ACM Trans. Archit. Code
  Optim.}, vol.~11, no.~4, Jan. 2015. [Online]. Available:
  \url{https://doi.org/10.1145/2687355}
\BIBentrySTDinterwordspacing

\bibitem{wholefuncv2011}
\BIBentryALTinterwordspacing
R.~Karrenberg and S.~Hack, ``{W}hole {F}unction {V}ectorization,'' in
  \emph{International Symposium on Code Generation and Optimization}, ser. CGO,
  2011. [Online]. Available:
  \url{http://www.cdl.uni-saarland.de/papers/karrenberg_wfv.pdf}
\BIBentrySTDinterwordspacing

\bibitem{pldi2018hack}
\BIBentryALTinterwordspacing
S.~Moll and S.~Hack, ``{P}artial {C}ontrol-flow {L}inearization,'' in
  \emph{Proceedings of the 39th ACM SIGPLAN Conference on Programming Language
  Design and Implementation}, ser. PLDI 2018.\hskip 1em plus 0.5em minus
  0.4em\relax New York, NY, USA: ACM, 2018, pp. 543--556. [Online]. Available:
  \url{http://doi.acm.org/10.1145/3192366.3192413}
\BIBentrySTDinterwordspacing

\bibitem{gpucheck}
\BIBentryALTinterwordspacing
T.~Lloyd, K.~Ali, and J.~N. Amaral, ``Gpucheck: Detecting cuda thread
  divergence with static analysis,'' Deparment of Computer Science, University
  of Alberta, Tech. Rep., 2019. [Online]. Available:
  \url{https://era.library.ualberta.ca/items/7ab2b28d-b111-448f-8273-2ff219132908}
\BIBentrySTDinterwordspacing

\bibitem{linearcfg2014}
J.~Anantpur and G.~R., ``Taming control divergence in gpus through control flow
  linearization,'' in \emph{Compiler Construction}, A.~Cohen, Ed.\hskip 1em
  plus 0.5em minus 0.4em\relax Berlin, Heidelberg: Springer Berlin Heidelberg,
  2014, pp. 133--153.

\bibitem{codemotion20}
J.~Fukuhara and M.~Takimoto, ``Branch divergence reduction based on code
  motion,'' \emph{Journal of Information Processing}, vol.~28, pp. 302--309,
  2020.

\bibitem{ccc2015}
\BIBentryALTinterwordspacing
F.~Khorasani, R.~Gupta, and L.~N. Bhuyan, ``Efficient warp execution in
  presence of divergence with collaborative context collection,'' in
  \emph{Proceedings of the 48th International Symposium on Microarchitecture},
  ser. MICRO-48.\hskip 1em plus 0.5em minus 0.4em\relax New York, NY, USA:
  Association for Computing Machinery, 2015, p. 204–215. [Online]. Available:
  \url{https://doi.org/10.1145/2830772.2830796}
\BIBentrySTDinterwordspacing

\bibitem{iterdelay2011}
\BIBentryALTinterwordspacing
T.~D. Han and T.~S. Abdelrahman, ``Reducing branch divergence in gpu
  programs,'' in \emph{Proceedings of the Fourth Workshop on General Purpose
  Processing on Graphics Processing Units}, ser. GPGPU-4.\hskip 1em plus 0.5em
  minus 0.4em\relax New York, NY, USA: Association for Computing Machinery,
  2011. [Online]. Available: \url{https://doi.org/10.1145/1964179.1964184}
\BIBentrySTDinterwordspacing

\bibitem{spec_reconv_cgo20}
\BIBentryALTinterwordspacing
S.~Damani, D.~R. Johnson, M.~Stephenson, S.~W. Keckler, E.~Yan, M.~McKeown, and
  O.~Giroux, ``Speculative reconvergence for improved simt efficiency,'' in
  \emph{Proceedings of the 18th ACM/IEEE International Symposium on Code
  Generation and Optimization}, ser. CGO 2020.\hskip 1em plus 0.5em minus
  0.4em\relax New York, NY, USA: Association for Computing Machinery, 2020, p.
  121–132. [Online]. Available: \url{https://doi.org/10.1145/3368826.3377911}
\BIBentrySTDinterwordspacing

\bibitem{common_subexpression_convergence}
S.~Damani and V.~Sarkar, ``Common subexpression convergence: A new code
  optimization for simt processors,'' in \emph{Languages and Compilers for
  Parallel Computing}, S.~Pande and V.~Sarkar, Eds.\hskip 1em plus 0.5em minus
  0.4em\relax Cham: Springer International Publishing, 2021, pp. 64--73.

\bibitem{tblockcompact}
W.~W.~L. {Fung} and T.~M. {Aamodt}, ``Thread block compaction for efficient
  simt control flow,'' in \emph{2011 IEEE 17th International Symposium on High
  Performance Computer Architecture}, 2011, pp. 25--36.

\bibitem{varwarp}
\BIBentryALTinterwordspacing
T.~G. Rogers, D.~R. Johnson, M.~O'Connor, and S.~W. Keckler, ``A variable warp
  size architecture,'' in \emph{Proceedings of the 42nd Annual International
  Symposium on Computer Architecture}, ser. ISCA '15.\hskip 1em plus 0.5em
  minus 0.4em\relax New York, NY, USA: Association for Computing Machinery,
  2015, p. 489–501. [Online]. Available:
  \url{https://doi.org/10.1145/2749469.2750410}
\BIBentrySTDinterwordspacing

\bibitem{warpsubdivision}
\BIBentryALTinterwordspacing
J.~Meng, D.~Tarjan, and K.~Skadron, ``Dynamic warp subdivision for integrated
  branch and memory divergence tolerance,'' \emph{SIGARCH Comput. Archit.
  News}, vol.~38, no.~3, p. 235–246, Jun. 2010. [Online]. Available:
  \url{https://doi.org/10.1145/1816038.1815992}
\BIBentrySTDinterwordspacing

\bibitem{multipath}
A.~{ElTantawy}, J.~W. {Ma}, M.~{O'Connor}, and T.~M. {Aamodt}, ``A scalable
  multi-path microarchitecture for efficient gpu control flow,'' in \emph{2014
  IEEE 20th International Symposium on High Performance Computer Architecture
  (HPCA)}, 2014, pp. 248--259.

\bibitem{ispc}
M.~Pharr and W.~R. Mark, ``ispc: A spmd compiler for high-performance cpu
  programming,'' in \emph{2012 Innovative Parallel Computing (InPar)}, 2012,
  pp. 1--13.

\bibitem{targeted_transformations}
\BIBentryALTinterwordspacing
C.~Cadar, ``Targeted program transformations for symbolic execution,'' in
  \emph{Proceedings of the 2015 10th Joint Meeting on Foundations of Software
  Engineering}, ser. ESEC/FSE 2015.\hskip 1em plus 0.5em minus 0.4em\relax New
  York, NY, USA: Association for Computing Machinery, 2015, p. 906–909.
  [Online]. Available: \url{https://doi.org/10.1145/2786805.2803205}
\BIBentrySTDinterwordspacing

\end{thebibliography}

\end{document}